\documentclass[a4paper,fleqn261,usenatbib]{mnras}

\usepackage{newtxtext,newtxmath}

\usepackage[T1]{fontenc}
\usepackage{ae,aecompl}

\usepackage{graphicx}	
\usepackage{amsmath}	
\usepackage{color}

\title{ Equation of State of Hot Dense Hyperonic Matter in the Quark-Meson-Coupling (QMC-A) model}

\author[J. R. Stone]{
J. R. Stone, $^{1,2}$
V. Dexheimer,$^{3}$
P.~A.~M.~Guichon,$^{4}$
A.~W.~Thomas,$^{5}$
S.~Typel$^{6}$
\\
$^{1} $Department of Physics and Astronomy, University of Tennessee, Knoxville, TN 37996, USA\\
$^{2}$Department of Physics (Astrophysics), University of Oxford, Keble Road OX1 3RH, Oxford, UK\\
$^{3}$Department of Physics, Kent State University, Kent, OH 44243 USA\\
$^{4}$DPhN, IRFU-CEA, Universit\'e Paris-Saclay, F91191 Gif sur Yvette, France\\
$^{5}$CSSM and CoEPP, School of Physical Sciences, University of Adelaide, Adelaide SA 5005, Australia\\
$^{6}$Fachbereich Physik,Institut für Kernphysik,Technische Universität Darmstadt, 64289 Darmstadt, Germany}

\date{Accepted XXX. Received YYY; in original form ZZZ}

\pubyear{2020}

\begin{document}

\label{firstpage}
\pagerange{\pageref{firstpage}--\pageref{lastpage}}
\maketitle

\begin{abstract}

We report a new equation of state (EoS) of cold and hot hyperonic matter constructed in the framework of the quark-meson-coupling (QMC-A) model. The QMC-A EoS yields results compatible with available nuclear physics constraints and astrophysical observations. It covers the range of temperatures from T=0 to 100 MeV, entropies per particle S/A between 0 and  6, lepton fractions from Y$_{\rm L}$=0.0 to 0.6, and baryon number densities n$_{\rm B}$=0.05-1.2 fm$^{\rm -3}$. Applications of the QMC-A EoS are made to cold neutron stars (NS) and to hot proto-neutron stars (PNS) in two scenarios, (i) lepton rich matter with trapped neutrinos (PNS-I) and (ii) deleptonized chemically equilibrated matter (PNS-II). We find that the QMC-A model predicts hyperons in amounts growing with increasing temperature and density, thus suggesting not only their presence in PNS but also, most likely, in NS merger remnants. The nucleon-hyperon phase transition is studied through the adiabatic index and the speed of sound c$_{\rm s}$. We observe that the lowering of (c$_{\rm s}$/c)$^{\rm 2}$ to and below the conformal limit of 1/3, is strongly  correlated with the onset of hyperons. Rigid rotation of cold and hot stars, their moments of inertia and Kepler frequencies are also explored. The QMC-A model results are compared with two relativistic models, the chiral mean field model (CMF), and the generalized relativistic density functional (GRDF) with DD2 (nucleon-only) and  DD2Y-T (full baryon octet) interactions. Similarities and differences are discussed.
 
\end{abstract}




\section{Introduction}

Properties of young proto-neutron stars (PNS) born in core-collapse supernovae (CCSN) have been one of the main topics of interest in observation and theoretical modeling for a long time (\cite{Burrows1986,Prakash1997,Pons1999}). Recently, the topic has resurfaced in the context of the possible emission of detectable gravitational waves (GW) in CCSN events (\cite{Ferrari2003,Camelio2017,Torres-Forne2019}). Better understanding of both static and dynamic properties of neutron stars (NS) will support an increasing interest in the emission of continuous GW from neutron star candidates in young supernova remnants \cite{Lindblom2020}. More generally, hot dense matter and its composition is particularly relevant in the context of binary neutron star mergers (BNSM) (\cite{Abbott2017,Baiotti2017,Radice:2016rys,Abbott2018,Most:2019onn,Bauswein2019}).

There has been an extensive discussion concerning the appearance of non-nucleonic species in dense matter, such as strange baryons (hyperons), pion and kaon condensates, and various  phases of quark matter, together with their density and temperature dependence (see e.g. \cite{Balberg1999,Pons1999,Pons2001,Mishra2010a,Chatterjee2016,Oertel2016,Oertel2017,Dexheimer2017,Roark2019,Malfatti2019}). 

The  effects of finite temperature and trapped neutrinos on properties of uniformly rotating PNS, using existing EOS of dense hot matter, consisting of nucleons and leptons (\cite{Lattimer1991}),  were first studied by \cite{Goussard1997}. They restricted themselves to several cases of chemically equilibrated matter at fixed temperature and/or entropy with/without fixed lepton fractions.

 \cite{Dexheimer2008a} constructed EoS of high density matter including baryon octet and decuplet in framework of the hadronic chiral SU(3) model,  applied to neutron and proto-neutron stars, taking into account trapped neutrinos, finite temperature, and entropy. The transition to the chirally restored phase was studied, and global properties of the stars such as minimum and maximum-mass configuration and radii were calculated for different cases. In addition, the effects of rotation on neutron star masses were investigated, as well as the conservation of baryon number and angular momentum to determine the maximum frequencies of rotation during different stages of stellar cooling.

A somewhat simplified variant of the QMC model (different from that used in the present work) has been used by \cite{Panda2010}  to explore neutrino-free matter and matter with trapped neutrinos in PNS at fixed temperatures up to 20 MeV and fixed entropy per particle S/A = 1 and  2. In comparison with the non-linear Walecka model with GM1 parameterization, QMC predicted smaller strangeness and neutrino fractions, however, growing faster with increasing temperature and density. 

Numerical simulations of BNSM both with nucleonic and hyperonic EoS with $\Lambda$ hyperons were performed by \cite{Sekiguchi2011}, who predicted a substantial role of hyperons in post-merger dynamics and suggested a possibility that the presence of hyperons may be imprinted in the evolution of the characteristic frequency of GW and the peak width of the GW signal.  

\cite{Burgio2011} constructed an EoS for hyperonic matter at finite temperature using the Bruckner-Hartree-Fock (BHF) method with the V18+UIX nucleon-nucleon and the NSC89 nucleon-hyperon interactions. Only $\Lambda$ and $\Sigma^-$ hyperons were predicted up to high densities when S/A was chosen to be 1 and 2. It was shown that the density threshold found for the appearance of hyperons at zero temperature does not exist at finite temperature and hyperons are present at all densities in the stellar core. The hyperonic content in matter with trapped neutrinos was found to be lower than in the neutrino free chemically-equilibrated matter. 

The impact of the hyperon-scalar-meson coupling on the EoS of hypernuclear matter at zero temperature and neutrino rich and neutrino free matter at finite temperature  was studied by \cite{Colucci2013} in the framework of the relativistic density functional theory with density-dependent couplings. The authors confirmed indirectly that the presence of neutrinos stiffen the EoS and dramatically change the composition of matter by keeping the fractions of charged leptons nearly independent of the density prior to the onset of neutrino transparency.

\cite{Raduta2020} used covariant density functional theory with the DDME2 interaction, extended to hypernuclear matter, and developed an EoS for matter in compact stars, including the full baryon octet and the $\Delta(1232)$ resonance, at non-zero temperatures. The calculation was done at constant values of entropy per particle, either assuming fixed lepton fraction (trapped neutrinos) or neutrinoless matter in chemical equilibrium. Universal relations between dynamical variables, such as the moment of inertia, quadrupole moment and tidal deformability, and the compactness of the star, as well as the I-Love-Q relation at non-zero temperature, were examined.

Several models were used to construct EoS tables, covering a wide range of temperatures, lepton fraction and a variety of compositions to be used in CCSN simulations. 

High-density hot matter, including nucleons, $\Lambda$, $\Sigma^{\rm 0,+,-}$ and $\Xi^{\rm 0,-}$ hyperons (the full baryon octet) and free thermal pions,
 was studied by \cite{Ishizuka2008} in the framework of an extended SU$_{\rm f}$(3) relativistic mean model (RMF). New EoS with hyperons, under the general label EOSY, were constructed and tables computed for a wide range of charge ratios, baryon densities and temperatures, mainly designed for CCSN simulations.

The EOSY tables were later adopted by \cite{Sumiyoshi2009} in their simulation of the dynamical collapse of a non-rotating massive stars. The exploration of the time dependence of the hyperon appearance in stages of the initial collapse, core bounce, and a temporary proto-neutron evolution before collapsing to a black hole revealed a significant hyperon content (mostly $\Lambda$ and $\Xi^-$) in the center of the star 680 ms after bounce, after being mostly located at about 10 km off center at 500 ms.  
 
\cite{Shen2011} constructed three EoS tables for CCSN simulations, assuming matter composed of nucleons and $\Lambda$ hyperons, in the framework of the RMF theory. They found that the population of $\Lambda$ hyperons increases with temperature; at T=100 MeV and the proton fraction Y$_{\rm p}$=0.1 it is distributed over the whole stellar-core, constituting  $\sim$16\% of the matter.

\cite{Oertel2012} modified the widely used Lattimer-Swesty EOS (\cite{Lattimer1991}) by including hyperons, pions and muons to the high density part of the EoS and showed that the additional degrees of freedom influence the thermodynamic properties such as pressure, energy density, and the speed of sound in a non-negligible way.

The EoS table including $\Lambda$ hyperons in dense matter, based on GRDF with the DD2 parameter set for the nucleons  (\cite{Typel2010}), and BHB$\Lambda$ and BHB$\Lambda\phi$ interactions to describe the  $\Lambda$ hyperons, was developed by \cite{Banik2014c}. Two variants of hyperonic EoS tables were constructed: in the np$\Lambda\phi$ case the repulsive hyperon-hyperon interaction mediated by the strange $\phi$ meson was taken into account, but in the np$\Lambda$ case it was not.

\cite{Marques2017} extended the DD2 interaction (\cite{Typel2010}) to the DD2Y version and used two scalar-isoscalar mesons fields, $\sigma$, coupling to all baryons and $\sigma^\ast$, coupling only to strange baryons. Two variants of the EoS, with the entire baryon octet, were constructed, without and with the $\sigma^\ast$ couplings, labeled DD2Y and DD2Y$\sigma^\ast$. The authors also examined hypermassive NS in the post-merger phase of BNSM and the moment of inertia-quadrupole moment (I-Q) universality. They found this universality was broken in fast rotating stars when thermal effects became important.

Two new general purpose EoS, applicable to neutron star mergers and CCSN simulations including the ful baryon octet, were developed by \cite{Fortin2018c}. The nucleonic EoS FSFHo \cite{Steiner2013} was extended to the SFHoY$^\ast$ EoS, imposing the SU(6) symmetry to obtain couplings of the isoscalar vector mesons, and the SFHoY EoS with the symmetry relaxed and empirical couplings used.

Despite the great variety of models and approaches in the literature, a general consensus on the EoS and composition of either cold or hot high density matter in the core of (proto)neutron stars, \textit{based on microphysics}, has not been achieved as yet (\cite{Stone2016}). The reason is that the nuclear and particle physics input, necessary to model the EoS, is poorly understood and there are no terrestrial data directly applicable to the high-density low-temperature sector of the QCD diagram (\cite{Sharma2019a}). 

The most recent trend in the field of study of EoS of high density matter in NS points toward statistical methods, such as Bayesian analysis (\cite{Naettilae2016,Raaijmakers2018,Lim2019,Greif2019}), parametric representations based on observational data (e.g. \cite{Oezel2016,Lindblom2018,Mena-Fernandez2019}), or machine learning methods (e.g. \cite{Fujimoto2018,Weih2019,Fujimoto2020}). However, this trend leaves many questions, related to the underlying quark structure of hadrons in dense matter, unanswered.

In this work we report a new microscopic EoS of hot high-density hyperonic matter using the latest version of the QMC model, called QMC-A, based on the QMC model detailed in (\cite{Guichon2018}), extended to finite temperature, as outlined in Sec.~\ref{sec:QMC-A}. The QMC-A model predictions of properties of static cold NS are presented in Secs.~\ref{sec:eos_cold} and \ref{sec:ns_cold} and of hot PNS in Sec.~\ref{sec:eos_hot} and \ref{sec:ns_hot}. These data are compared with the outcome of the chiral mean field model (CMF) (\cite{Dexheimer2008a,Roark2019}), and GRDF-Y model with the DD2Y-T interaction (\cite{Typel2020}). The nucleon-hyperon phase transition and its consequence for the stability of stars, together with the speed of sound in both cold and hot stars, are also discussed in these sections. Effects of the uniform rotation on the stellar masses, radii and composition, as well as universal relations between the moment of inertia and the stellar mass and compactness, are illustrated in Sec.~\ref{sec:rotation}. Discussion of the results and outlook can be found in Sec.~\ref{sec:discussion}.

\section{The Method}
\label{sec:method}

\subsection{The QMC-A model}
\label{sec:QMC-A}

The QMC model is an effective relativistic mean field model, different from other mean-field microscopic models used up to now.  In standard RMF-like mean models, the interaction between baryons is mediated by exchange of virtual mesons between point-like particles with no internal structure. In the QMC framework (\cite{Guichon1988,Guichon1996}), this interaction  takes place \textit{self-consistently} between valence quarks, confined in non-overlapping baryons. The effect of dense medium surrounding the baryons such as, for example, inside NS cores, on their interaction, is modeled by dynamics of the quarks inside the individual particles. The first version of the relativistic QMC model applied to NS (\cite{RikovskaStone:2006ta}) predicted the existence of cold NS with $\Lambda$ and $\Xi^{\rm 0}$ hyperons in their cores with a maximum mass of 1.97 M$_\odot$, three years before such a star was observed by \cite{Demorest:2010} (later updated by \cite{Fonseca2016} and \cite{Arzoumanian2018a}). The non-relativistic application of the QMC model to finite nuclei yielded predictions of ground state properties of finite nuclei in excellent agreement with experimental data (\cite{Stone2016a,Martinez2019,Stone2019,Martinez2020}).

In the QMC model, the parameters of the quark bag model, representing baryons, are adjusted to reproduce the masses of the baryon octet in free space. The meson fields are coupled to the quarks with coupling constants which are, for convenience, expressed in terms of the nucleon-meson couplings in free space  $g_{\sigma  N}$, $g_{\omega N}$ and $g_{\rho  N}$. The numerical values of the couplings are adjusted to reproduce the properties of symmetric nuclear matter.

These couplings, together with the mass of the $\sigma$ meson $M_\sigma$ (not well defined experimentally) and the radius of the bag (here chosen to be 1 fm to reproduce the radius of the proton), form a a unique set of five parameters. Once determined, the set is fixed and cannot be varied to improve the predictive power of the model. Should a discrepancy between the model prediction and new observational and experimental data occur, missing physics in the model must be sought. The numerical values of the parameters used in this work can be found in Table~\ref{tab:1}. It is important to note that in the nuclear medium these couplings acquire an effective density dependence which is determined by the response of the quark structure of the baryons to the meson fields. This is to be contrasted with other models, such as DD2Y, where the density dependence is added by hand.

Because in the QMC model the forces are acting between valence quarks inside baryons and not between point-like baryons without internal structure, there is no need to increase the number of parameters when the baryonic composition of matter changes. In other words, the matter consisting of only nucleons or of the entire baryon octet, is modeled by the same parameter set. As a consequence of the unique features of the QMC model, the hyperon-nucleon and hyperon-hyperon couplings are \textit{fixed  by the quark structure } in free space as well as in dense matter. This is  contrary to most conventional  RMF-like models with hyperons, where these  couplings are fitted or have to be determined by symmetries.

In QMC-A, the formalism has been extended to include matter at finite temperatures with and without neutrinos and a smooth transition between T=0 and T>0 cases has been achieved. The model includes a new treatment of the $\sigma$-field and a the first complete, self-consistent treatment of the Fock terms, which will be described elsewhere. The QMC-A EoS complies with the commonly accepted constraints on the isospin-symmetric nuclear matter at saturation saturation density $n_{\rm 0}$, saturation energy $E_{\rm sat}/A$, the symmetry energy $J$, the slope of the symmetry energy $L$, and incompressibility $K$ (\cite{Tsang2012,Horowitz2014, Stone2014a}), as demonstrated in Table~\ref{tab:1}. 

\subsection{Calculation details}
\label{sec:calc}

All calculations in this work have been performed assuming a full chemical and thermal equilibrium, and the charge neutrality being strictly conserved. For a system consisting of the octet baryons and leptons $(e^{-},\bar{\nu}_{\rm e},e^{+},\nu_{\rm e}),(\mu^{-},\bar{\nu}_{\mu},\mu^{+},\nu_{\mu})$, we calculate chemical potentials of all the constituents, consequently used to derive other thermodynamical quantities. We assume equilibrium for strangeness changing weak interactions and take strangeness chemical potential equal to zero throughout the work (\cite{Oertel2012}) so that, for example, $\mu_n=\mu_\Lambda$ and $\Sigma^-=\mu_n+\mu_e$. Further assuming lepton number conservation, a system with the baryon number density $n_{\rm B}$  is described by the lepton fraction $Y_{\rm L}$= ($L_{\rm e}$+  $L_{\mu})$/$n_{\rm B}$ where electron, $L_{\rm e}$, and muon,  $L_{\mu}$,  lepton number densities are supposed to be known and enter the equilibrium and charge conservation relations
\begin{eqnarray}
\mu(e^{-})-\mu(\nu_{e}) & = & \mu(\mu^{-})-\mu(\nu_{\mu})\ ,\\
\mu(i) & = &\mu(n) - Q(i) \mu(e^{-})+Q(i) \mu(\nu_{e})\ ,
\end{eqnarray}
and 
\begin{eqnarray}
n(e^{-})+n(\nu_{e})-n(e^{+})-n(\bar{\nu}_{e}) & = & L_{e}\ ,\\
n(\mu^{-})+n(\nu_{\mu})-n(\mu^{+})-n(\bar{\nu}_{\mu}) & = & L_{\mu}\ ,\\
n(e^{+})-n(e^{-})+n(\mu^{+})-n(\mu^{-}) & = & -\sum_{i}n(i)Q(i)\ ,\\
\sum_{i}n(i) & = & n_{B}\ ,
\end{eqnarray}
where $n(i)$, $\mu(i)$ and $Q(i)$ are particle number densities, chemical potentials and charges of a baryon constituent $i$, respectively. This scenario corresponds to dense matter with trapped neutrinos, believed to exist, at least to certain extend, between 0.1 to 1.0 s after the bounce in the CCSN event (\cite{Prakash1997,Pons1999,Camelio2017}). The system can be modeled at fixed values of Y$_{\rm L}$ and temperature/entropy per particle. Although in the dynamic development of a PNS during this regime, the chemical equilibrium is not strictly speaking observed for leptons, this is the approximation we take. As we do not have neutrino transport built in the model to inform us about the changing number of neutrinos in the system, $Y_{\rm \nu}$ cannot be self-consistently determined.

 During the neutrino diffusion from the core to the surface, matter becomes deleptonized through several processes, mainly neutrino-nucleon absorption (\cite{Prakash1997}), loss of electron neutrinos (\cite{Pons1999}) and loss of electrons through electron capture (\cite{Burrows1981}). \cite{Janka1995} noted that, although electron-positron annihilation, nucleon-nucleon bremsstrahlung, and even neutrino-antineutrino annihilation contribute non-negligibly to the rate of deleptonization of the entire PNS, the reduction of the electron chemical potential to close to zero in the core (i.e. not the shocked mantle of the PNS) is dominated by electron capture.  \cite{Fischer2016} provided a list of medium modified charge- and neutral current weak processes, potentially participating in the deleptonization process of the PNS, including elastic scattering of neutrinos on nucleons and inelastic scattering on electron/positrons and pair processes. The deleptonization  processes produce heating and the entropy of the star grows. 

Deleptonized matter in chemical equilibrium is governed by equations
\begin{eqnarray}
\mu(e^{-}) & = & \mu(\mu^{-})\ ,\\
\mu(i) & = & \mu(n) -Q(i) \mu(e^{-})\ ,
\end{eqnarray}
and
\begin{eqnarray}
n(e^{+})-n(e^{-})+n(\mu^{+})-n(\mu^{-}) & = & -\sum_{i}n(i)Q(i)\ ,\\
\sum_{i}n(i) & = & n_{B}\ .
\end{eqnarray}
  
In this work, we adopt two illustrative scenarios. In the first, (PNS-I), we consider matter with lepton fraction Y$_{\rm L}$=0.4, trapped neutrinos and the entropy per particle S/A=1. In the second, PNS-II, the neutrinoless chemically equilibrated matter (NS($\beta$)), developed after the deleptonization process has finished, has S/A=2. Note that we omit the Boltzmann constant $k_{\rm B}$ because we are using natural units, in which case it assumes the value one.

Admittedly, both scenarios provide a rather schematic picture of the PNS birth and development. In modern CCSN simulations, all thermodynamical variables of the PNS, such as temperature, $Y_{\rm L}$ and $S/A$, depend on the model of the progenitor star, details of the collapse, and development of the shock,  which are themselves locally dependent on the density and composition distribution of the core material, determined by a chosen EoS (\cite{Hix2003b,Lentz2012}). However, we believe that the two examples described above provide an illustrative trend.

To facilitate better understanding of the role of microphysics in the EoS of cold and hot compact objects and to provide a comparison with other models, the Chiral-Mean-Field (CMF) model and the generalized relativistic density functional (GRDF) with the DD2 (nucleon-only) and the  DD2Y-T (full baryon octet) interactions, were chosen.

 The Chiral-Mean-Field (CMF) model is based on a nonlinear realization of the SU(3) sigma model (\cite{Papazoglou:1998vr}). It is an effective quantum relativistic model, which naturally fulfills the causality limit, and
describes hadrons and quarks interacting via meson exchange ($\omega$, $\sigma$, $\rho$, $\delta$, $\phi$, and $\zeta$). It is constructed in a chirally invariant manner, with particle masses originating from interactions with the medium and, therefore, decreasing at high densities and/or temperatures. The model is in agreement with standard nuclear and astrophysical constraints (\cite{Dexheimer2008a,Roark2019,Dexheimer:2018dhb}), as well as lattice QCD and perturbative QCD (\cite{Dexheimer:2009hi,Roark2018}). The hyperon couplings are determined following SU(3) and SU(6) coupling schemes, with the scalar sector being chosen to reproduce the hyperon vacuum masses. The only free coupling left is for the $\Lambda$ hyperon, which is fitted in order to reproduce the expected value  $\Lambda$ single-particle potential at saturation $U_{N\Lambda}$ $(n_{0}$).

The DD2Y-T model (\cite{Typel2020}) is an extension of a relativistic energy density functional  with $\omega$, $\sigma$, and $\rho$ mesons by including the entire baryon octet and the $\phi$ meson. The meson-nucleon couplings are assumed to depend on the total baryon density using the parametrization DD2 (\cite{Typel2010}).  As in the DD2Y model \cite{Marques2017}, the couplings of the hyperons to the scalar $\sigma$, and the vector $\omega$ and $\rho$ mesons are density dependent. The ratios of hyperon to nucleon couplings are constant and the couplings for $\omega$ and $\rho$ mesons are determined by SU(6) symmetry. The $\sigma$ meson-hyperon coupling is chosen by fixing the values of the single-particle hyperonic potentials in nuclear matter at saturation density. But, in the DD2Y-T model, the $\phi$ meson-hyperon couplings ($\phi$ does not couple to nucleons), are constant and are determined at saturation from the $\omega$ couplings, with the corresponding SU(6) factors for different hyperons. There is a also minor difference in the $\Sigma$ hyperon masses; \cite{Marques2017} adopt a common mass for all $\Sigma ^{(0,+,-)}$ and in the DD2Y-T model their experimental masses are used.  Basic parameters of the QMC-A, CMF and DD2Y-T models are summarized in Table~\ref{tab:1}.

\begin{table}
\footnotesize
\setlength{\tabcolsep}{0pt}
\centering
\caption{\label{tab:1} Coupling constants, meson rest masses, single-particle potentials and the symmetric nuclear matter properties at the saturation  density  $n_{\rm 0}$ as computed in the QMC-A, CMF and DD2 models. The single-particle potentials are \textit{calculated} in the QMC-A model but are taken as parameters in the CMF and DD2 models (see Sec.~\ref{sec:calc}). The meson-nucleon couplings in the DD2 model given here are the saturation density. Data are taken from Ref. \protect\cite{Roark2018,Typel2010,Typel2020} and the current calculation. In the CMF model, as in other chiral models, $\sigma$ is the chiral condensate, not necessarily related to the physical $\sigma$ meson.}
\vspace{5pt}
  \begin{tabular}{ p{4cm}p{1.5cm}p{1.5cm}p{1.5cm} }
 Model                                       &       QMC-A   &  CMF &	DD2Y-T\\ \hline   
   g$_{\rm\sigma N}$          &     11.40          &  -9.83      &    10.68        \\
   g$_{\rm \omega N}$         &     9.73         &  11.9       &     13.34        \\
   g$_{\rm \rho N}$              &    6.70        &    4.03        &        3.63        \\
    M$_\sigma$ [MeV]                   &     700          &    $-$        &   546       \\ 
     M$_\omega$ [MeV]                  &      783         &   781         &   783       \\               
    M$_\rho$ [MeV]                      &       775         &   761         &       763       \\   
   U$_{\rm N\Lambda}$ (n$_{\rm 0}$)  [MeV]         &        -28       &    -28   &   -28       \\
    U$_{\rm N\Sigma}$ (n$_{\rm 0}$)  [MeV]         &       -0.96      &    +5.3  &  +30      \\
     U$_{\rm N\Xi}$ (n$_{\rm 0}$)     [MeV]          &       -12.7      & -18.7    &   -14      \\   \hline
      n$_{\rm 0}$ [fm$^{\rm -3}$]                  &       0.156      &     0.15        & 0.15    \\
      E$_{\rm sat}$/A [MeV]                 &     -16.2        &   -16          &  -16.2 \\
       J   [MeV]                             &     28.5          &     30         & 32.7  \\
      L   [MeV]                             &     54             &    88          &  58     \\
      K   [MeV]                             &     292           &    300         &   243  \\  \hline

\end{tabular}
\end{table}

 The DD2 model with light and heavy clusters, as described in (\cite{Pais2017}), was used in all three models for inhomogeneous matter at sub-saturation densities at T=0 MeV case, as only in this case such matter exists in NS. Here the DD2 calculation gives a fully thermodynamically consistent result, with the crust including phase transitions between different forms of nuclear and particle species, and the transition from inhomogeneous to homogeneous matter at higher densities. In the DD2Y-T model at T=0,  hyperons appear only at densities somewhat above the saturation density. Below the lowest hyperon density threshold, the EoS is identical to the DD2 model. Thus the inhomogeneous matter is not affected by the presence of hyperons and everything is again thermodynamically consistent. At (sufficiently high) finite temperature the clusters dissolve and there is no density threshold for appearance of hyperons. The DD2Y-T EoS is again thermodynamically consistent for all densities.

The transition between homogeneous and inhomogeneous matter was treated individually in each case, searching for the baryon number density region where there is a simultaneous smooth connection between the energy density, pressure and particle number density and their derivatives, and interpolating over this region if necessary.

The new QMC-A EoS covers the range of temperatures from T=0 to 100 MeV, entropies per particle S/A between 0 to 6, lepton fractions from Y$_{\rm L}$=0.0 to 0.6, and baryon number density range n$_{\rm B}$=0.05 - 1.2 fm$^{\rm -3}$. It also provides neutrino energies in the scenario PNS-I. The QMC-A EoS can be used in CCSN and NS merger simulations and extended beyond the current range if necessary. 

\section{Results}
\label{sec:results}

\subsection{The Equation of state and composition of high density matter}
\subsubsection{Cold matter}
\label{sec:eos_cold}

\begin{figure}
  \includegraphics[trim={1.5cm 2.5cm 1.5cm 2.7cm},width=9.0cm]{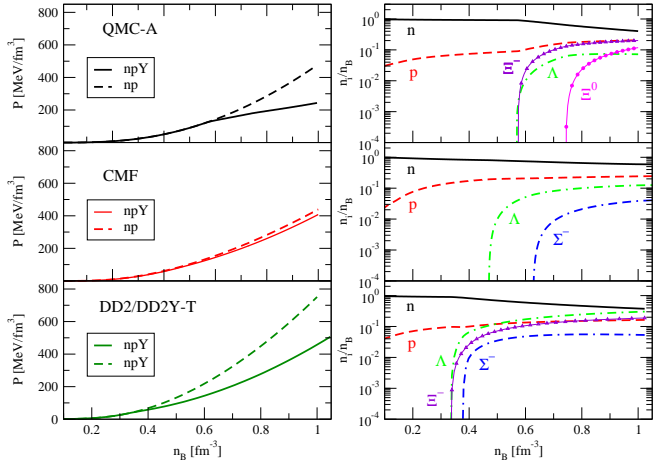}
  \caption{Equations of state of the QMC, CMF, DD2 (for nucleons only) and DD2Y-T (for nucleons and hyperons) models at T=0 MeV. Pressure vs. baryon number density in matter with the full baryon octet (npY) and nucleonic (np) matter without hyperons is shown in the left panels. Relative population of nucleons and hyperons in units of the total baryon number density $n_{\rm B}$ are displayed in the right panels. Only population fractions higher than 10$^{\rm -4}$ in the region of n$_{\rm B}$ between  0.1 - 1 fm$^{\rm -3}$ are shown.}
  \label{fig:1}
\end{figure}

We start with applying the QMC-A model to construct an EoS for the core of cold NS, assuming charge-neutral, chemically equilibrated homogeneous matter, containing the full baryon octet, electrons and muons. In the left panels of  Fig.~\ref{fig:1}, we plot the pressure dependence on the baryon number density for matter containing the entire baryon octet for the three models QMC-A, CMF, and DD2Y-T, with the last two being used for a comparison.  The density range is limited to 0.1 - 1.0 fm$^{\rm -3}$, as the central densities of most cold NS models are expected to lie in this region. As anticipated, the presence of hyperons softens the EoS, leading to lowering the slope of the pressure with the onset of hyperons, when compared to a pure nucleonic EoS.  Different patterns of the softening are clearly related to the density distribution of the hyperon population, shown in the right panels of Fig.~\ref{fig:1} for particle fractions higher than 10$^{\rm -4}$. One can easily conclude that the onset densities and the amount of individual species are clearly model dependent.
 
It is interesting to note that \cite{Abbott2018} reported a constraint on pressure at twice the nuclear saturation density to be  between 11.23 and 38.7 Mev/fm$^{\rm 3}$ (with 90\% confidence), derived from the analysis the GW170817 signal. Hyperons are not predicted to be present at this density in this work. For matter composed of only nucleons, all three models predict very close values of pressure 32.99, 32.93 and 32.14  Mev/fm$^{\rm 3}$ for QMC-A, CMF and DD2Y-T, respectively, well within the limits extracted from the GW observation.

The hyperons appear naturally at T=0 in nucleonic matter (due to the Pauli blocking), when their chemical potentials, increasing with density of the degenerate matter, reach values above that of their effective masses. Eventually, strangeness non-conserving weak processes become possible, creating the hyperon population in the stellar core (\cite{Glendenning1985a,Balberg1999,Glendenning2012}). The variation in the threshold densities is related to the differences in hyperon couplings and the consequent hyperon binding energies, defining their chemical potentials, in different models (see e.g. recent study by \cite{Fortin2020}).   

Examination of the right panels of Fig.~\ref{fig:1} reveals that $\Lambda$ hyperons are predicted to appear first, at threshold densities 0.5 - 0.6, 0.4 - 0.5 and 0.3 - 0.4 fm$^{\rm -3}$, in QMC-A, CMF and DD2Y-T models, respectively. The other members of the full hyperon octet, included in the models, appear at different threshold densities. The QMC-A model predicts only $\Xi^{\rm 0,-}$ hyperons above 0.5 fm$^{\rm -3}$ , CMF shows just the $\Sigma^-$ hyperon appearing at densities above 0.6 fm$^{\rm -3}$ and, in the DD2Y-T model, both negatively charged hyperons, $\Sigma^-$ and $\Xi^{-}$, are present at rather low densities below 0.4 fm$^{\rm -3}$. 

In the QMC-A model, $\Sigma$ hyperons do not appear at baryon number densities below n$_{\rm B}$=1.0 fm$^{\rm -3}$. This effect was recognized already in our early work (\cite{RikovskaStone:2006ta,Guichon2018}). As discussed in Sec.~\ref{sec:QMC-A} in the QMC model,  the nucleon-hyperon and hyperon-hyperon interactions are not a subject of choice, but emerge naturally from the formalism. In particular, the hyperfine interaction that splits the $\Lambda$ and $\Sigma$ masses in free space is significantly enhanced in-medium (\cite{Guichon:2008zz}), leading to what is effectively a repulsive three- body force for the $\Sigma$ hyperons, with no additional parameters. The absence of $\Sigma$ hyperons in cold matter is supported by the fact that no bound $\Sigma^-$ hypernuclei at medium or high mass has been found as yet, despite dedicated search (\cite{Harada2006g,Harada2015o}). However, the CMF model predicts a considerable presence of $\Sigma^-$ at about four times the nuclear saturation density and in the DD2Y-T model the $\Sigma^-$ appear in the density region below 0.4 fm $^{\rm -3}$.

The $\Xi^{(\rm 0,-)}$ hyperons, not predicted by the CMF model at densities below 1.0 fm$^{\rm -3}$, appear in the QMC-A model at density almost identical to the threshold for $\Lambda$. Only the negative $\Xi^{-}$ is observed in the DD2Y-T model. The presence of the $\Xi$ hyperons at rather low densities indicates an attractive nucleon-$\Xi^{(\rm 0,-)}$ force and points to existence of bound $\Xi^-$ hypernuclei. So far, two single events involving $\Xi$ hypernuclei, $^{\rm 12}_{\Xi^-}$Be (\cite{Kchaustov2000}) and $^{\rm 15}_{\Xi^-}$C (\cite{Nakazawa2015d}), have been reported (\cite{Yoshida2019}). 

An important thermodynamic quantity, closely related to the EoS, is the adiabatic index $\Gamma$, a sensitive indicator of phase changes in stellar matter and the stability with respect to vibrations of a star (\cite{Akmal1998,Haensel2002,Chamel2008,Casali2010} ). It is defined as
\begin{equation}
\Gamma = \frac{d \log P}{d \log n_B}=\frac{n_B}{P}\frac{d P}{d n_B}.
\end{equation}
Polytropic EoS have constant $\Gamma$, equal to 4/3 (5/3) for a relativistic (non-relativistic) free-Fermi gas. Using realistic nucleonic EoS, (\cite{Chamel2008}) studied supernova matter with trapped neutrinos and Y$_{\rm L}$=0.4 and S/A=1 and found that $\Gamma$ was continuously growing with density from 0.5 to 4. For multicomponent matter, $\Gamma$ exhibits jumps at densities coincident with density thresholds of individual components, signaling phase transitions and/or changes in the make-up of the matter.

\begin{figure}
  \includegraphics[trim={0 2.5cm 0 1.0cm},width=9.0cm]{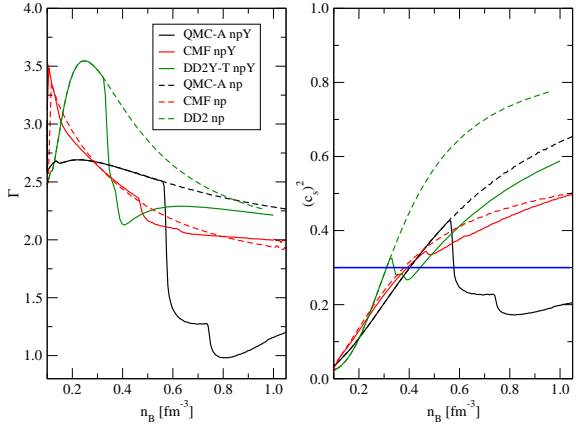}
  \caption{The adiabatic index (left) and the square of the speed of sound in units of c$^{\rm 2}$ (right) vs. baryon number density, as computed in the QMC, CMF and DD2/DD2Y-T models at T=0 MeV. Results for matter containing the full baryon octet (npY) and nucleonic matter without hyperons (np) are shown. The blue horizontal line indicates the conformal limit $c^{\rm 2}_{\rm s}$=1/3.}
  \label{fig:2}
\end{figure}

 We present in the left panel of Fig.~\ref{fig:2} the adiabatic index $\Gamma$ as calculated in the QMC-A, CMF and DD2/DD2Y-T models, in the range of density from 0.1 to 1 fm$^{\rm -3}$ (0.625 - 6.25 n$_{\rm 0}$), predicted to be reached in NS cores by realistic models. We see significant drops in the values of $\Gamma$ at densities which are coincident with the threshold densities for appearance of hyperons (see Fig.\ref{fig:1}). The large drop at density 0.5-0.6 fm$^{\rm -3}$ for QMC-A takes place because the $\Lambda$ and $\Xi^-$ hyperons appear at almost the same density. The next drop is clearly related to appearance of the $\Xi^{\rm 0}$. The CMF model shows much smaller drops at the threshold densities for the $\Lambda^-$ and $\Sigma^-$ hyperons as the amount of hyperons is smaller. The DD2Y-T model predicts very close density thresholds for the $\Lambda^-$ , $\Sigma^-$ and $\Xi^-$ hyperons, which can be associated with the large unresolved drop below 0.4 fm$^{\rm -3}$.  These drops are manifestations of instabilities in hyperonic matter, leading to vibrations which may be damped by various processes, including bulk viscosity, before reaching equilibrium (\cite{Jones2001,Haensel2002,Lindblom2002}). 

The adiabatic index is closely related to the speed of sound in units of $c$ (here equal to 1)
\begin{equation}
c_{\rm s}^{\rm 2}=\frac{dP}{d\epsilon} ,
\end{equation}
where $\epsilon$ is the energy density. The stability and causality conditions set limits 0$\leq c^{\rm 2}_{\rm s}\leq$ 1. Recent limits on the speed of sound derived from astrophysical observation were obtained by \cite{Bedaque2015} using a parameterized EoS and by \cite{Tews2018}  in the framework of the chiral effective field theory. It has been argued that the speed of sound in NS matter exceeds the \textit{conformal limit} (\cite{Cherman2009}) $c^{\rm 2}_{\rm s} \leq$ 1/3. All these studies were applied either to rather low densities or did not explicitly include hyperonic degrees of freedom. Very recently, \cite{Annala2020}, using their pQCD model extrapolated to densities relevant to NS, suggested that compliance with the conformal limit signals sizable quark-matter core in massive NS. The density dependence of $c^{\rm 2}_{\rm s}$ reported by \cite{Annala2020}, depicting the phase transition from hadronic to quark matter, differs from that for a transition from nucleonic to hyperonic matter both in nature and strength. However, in both cases, the transition region gives rise to $c^{\rm 2}_{\rm s} \leq$ 1/3. Recent discussion on the impact of irregularities in speed of sound in dense matter on the macroscopic neutron-star properties  (\cite{Tan2020}) inspires more investigation along these lines.

As is demonstrated in the right panel of Fig.~\ref{fig:2}, the QMC-A model predicts the  $c^{\rm 2}_{\rm s}$ lower than 1/3 at densities above ~0.6 fm$^{\rm -3}$. The CMF and DD2Y-T models yield $c^{\rm 2}_{\rm s}$ around 1/3 at densities between 0.3-0.5 fm$^{-3}$. It is not surprising that the instabilities, seen in the adiabatic index, reflect also in the density dependence of the speed of sound. In a classical analog, induced vibrations in the medium interfere with propagation of the sound wave, thus causing its impedance through the refractive index. Because the instability is larger in the QMC-A model than in the other models, the effect on the speed on the sound is more pronounced.  Detailed analysis of these conjectures goes beyond the scope of this paper and will be addressed separately. However, we can rule out with certainty that the conformal behaviour of $c^{\rm 2}_{\rm s}$ is a unique signature of a phase transition to quark matter in the NS core. None of the models used here includes quark degrees of freedom and yet the speed of sound is predicted to be below or close to the conformal limit. As the QMC-A, CMF and DD2Y-T EoS with nucleons and hyperons are in full compliance with all known astrophysical constraints, $\Gamma$ and $c^{\rm 2}_{\rm s}$ naturally also reflect these constraints.

\subsubsection{Matter at finite temperature}
\label{sec:eos_hot}

\begin{figure}
  \includegraphics[trim={0 2.5cm 0 1.0cm},width=9.0cm]{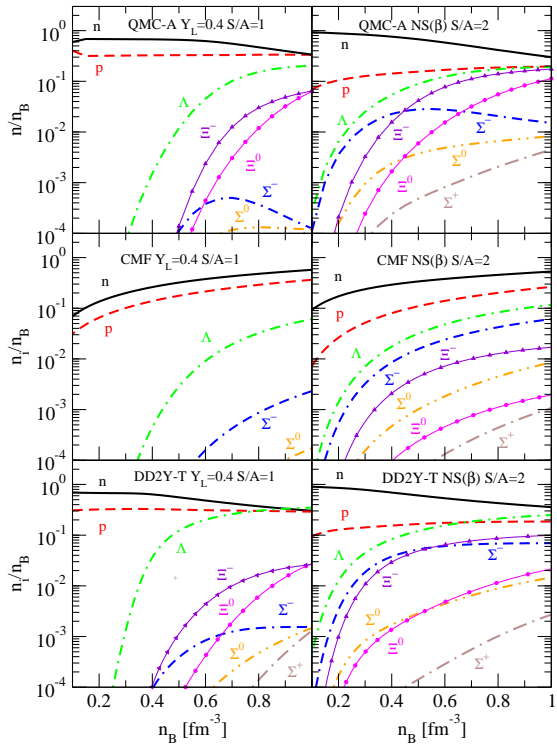}
  \caption{The same as the right panels in Fig.~\ref{fig:1} but for the PNS-I (left)  and PNS-II (right) scenarios. Only population fractions higher than 10$^{\rm -4}$ in the region of the baryon number density between 0.1 - 1 fm$^{\rm -3}$ are shown.}
  \label{fig:3}
\end{figure}

\begin{figure}
 \includegraphics[trim={0 2.5cm 0 1.0cm},width=9.0cm]{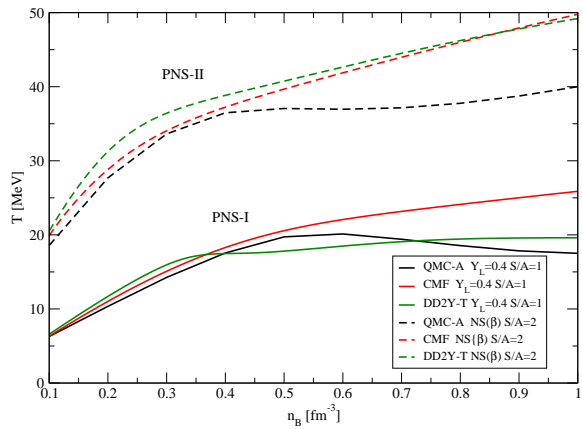}
  \caption{Temperature distribution vs. baryon number density in hyperonic matter in the PNS-I (solid curves) and PNS-II (dashed curves) scenarios as calculated in the QMC-A, CMF and DD2Y-T models.}
  \label{fig:4}
\end{figure}

We have demonstrated in the previous section that the number density dependence of the thresholds for the appearance of hyperons in dense matter and their population in cold stars are model dependent. This model dependence is still apparent at finite temperatures, where the temperature effects smear out the density thresholds. As a consequence, hyperons appear at low-densities, populating a large portion of cores of warm low-mass NS. The population distribution of nucleons and hyperons at the two scenarios, PNS-I and PNS-II, adopted in this work, are presented in Fig.~\ref{fig:3}. Comparing Fig.~\ref{fig:1} and Fig.~\ref{fig:3} indicates an increase of hyperonic content in the PNS-I, followed by a more dramatic increase in the scenario PNS-II as compared to the T=0 MeV case. In the latter, population of the entire octet in fractions larger than 10$^{-4}$ is predicted by all three models at densities below the respective thresholds at T=0 MeV. Our results are consistent with findings of other studies (see e.g. \cite{Prakash1997,Burgio2011,Rabhi2011,Oertel2016,Marques2017}). Matter with trapped neutrinos is lepton rich with an increased proton fraction (and a lower the neutron fraction) at low densities, thus increasing the hadronic electric charge Y$_{\rm Q}$. The onset of hyperons is in this case shifted to higher densities (compare e.g. Figure 4 (panels e and f) in \cite{Rabhi2011}). The amount of negatively charged hyperons increases with decreasing Y$_{\rm e}$ during the deleptonization of a hot PNS in order to maintain charge neutrality. At the same time, the EoS softens at higher densities leading to a lower maximum mass of a lepton-poor star. The temperature distribution as a function of baryon number density, $n_{\rm B}$,  is illustrated in Fig.~\ref{fig:4}, demonstrating the predicted rise in temperature in deleptonized matter (scenario PNS-II) as compared to the matter with trapped neutrinos (scenario PNS-I).

\begin{figure}
  \includegraphics[trim={0 8.5cm 0 1.0cm},width=9.5cm]{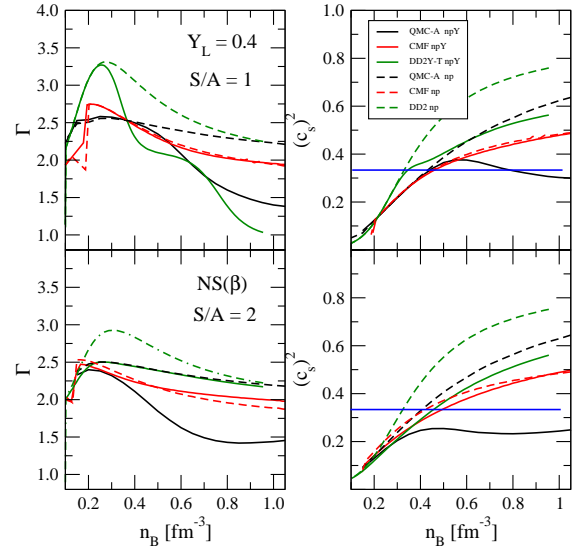}
  \caption{The same as Fig.~\ref{fig:2} but for the PNS-I (top panels) and PNS-II (bottom panels) cases.}
  \label{fig:5}
\end{figure}

The temperature effects on the adiabatic index and the speed of sound at finite temperature is illustrated in Fig.~\ref{fig:5}. We observe a much smoother density dependence of $\Gamma$, reflecting the absence of the density thresholds for appearance of hyperons, present in the T=0 MeV case. In the QMC-A model, $c^{\rm 2}_{\rm s}$ is almost constant below 1/3 in the PNS-II scenario. In the PNS-I case, $c^{\rm 2}_{\rm s}$ remains below 0.4, again demonstrating sensitivity to changes in the hyperon population which increases with growing density, as demonstrated in Fig.~\ref{fig:3}. In the CMF and DD2Y-T models, there is very little difference between the density dependence of  $c^{\rm 2}_{\rm s}$ in the PNS-I and PNS-II cases, showing a smooth growth up  $c^{\rm 2}_{\rm s}$ to the value of about 0.6 at density 1.0 fm$^{\rm -3}$.

\subsection{Neutron star masses and radii}
\label{sec:ns}
\subsubsection{Cold neutron stars}
\label{sec:ns_cold}

\begin{figure}
  \includegraphics[trim={0 2.5cm 0 1.0cm},width=9.0cm]{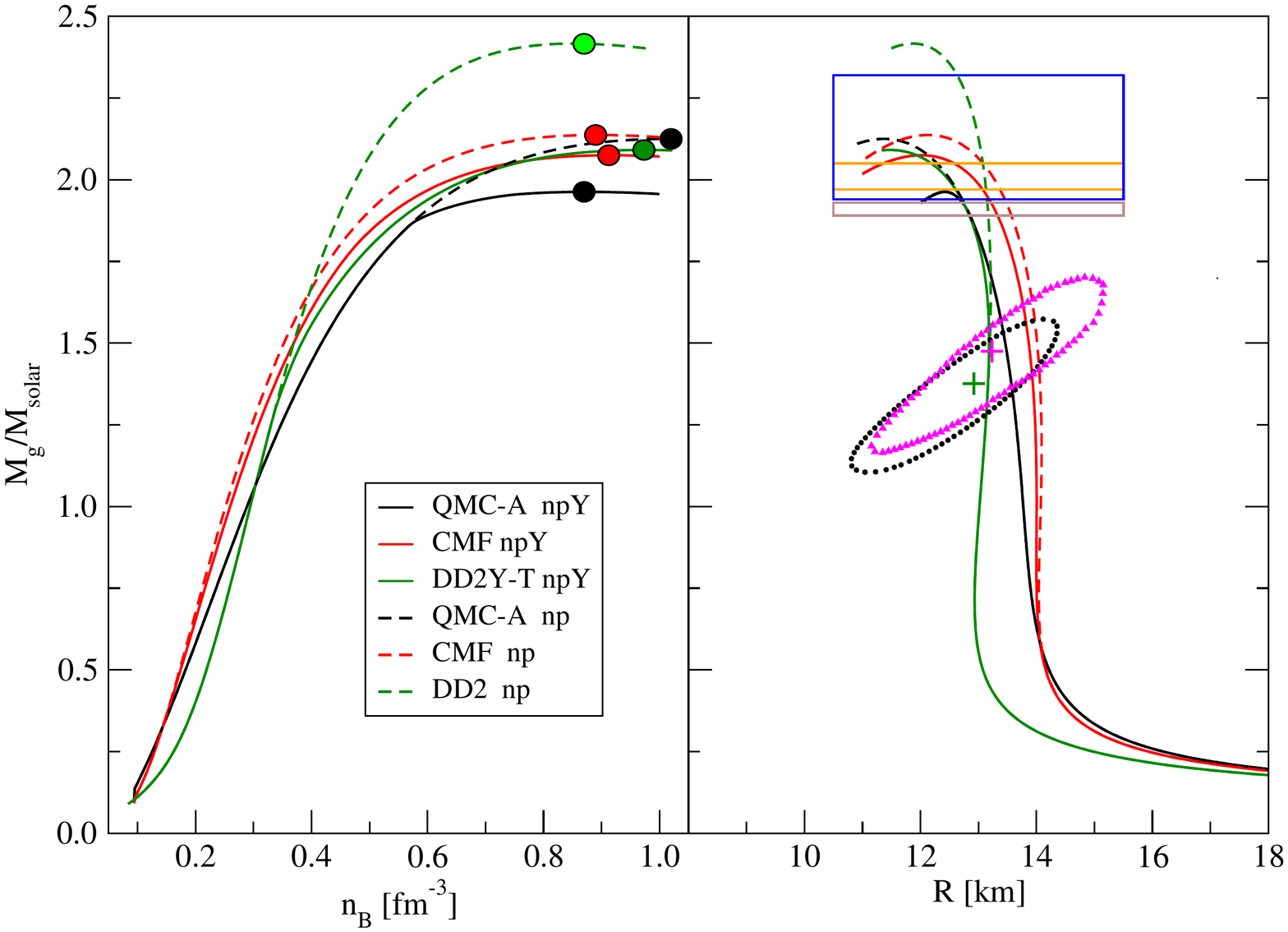}
  \caption{Neutron star gravitational masses vs. the central baryon density (left) and radius (right), computed with the QMC-A, CMF and DD2Y-T models for hyperonic (solid line) and nucleonic (dashed) matter at T=0 MeV. Observational limits on the maximum-mass configuration and its radius are illustrated by the blue (\protect\cite{Cromartie2019}), orange (\protect\cite{Antoniadis:2013pzd}) and brown (\protect\cite{Arzoumanian2018a})  solid rectangles. Recent data from NICER, analysed independently by \protect\cite{Miller2019} (magenta) and by \protect\cite{Riley2019} (dark green) yielded limits on the gravitational mass and radius of PSR J0030-0451. M-R contours enclosing 68\% of the posterior mass as obtained by \protect\cite{Miller2019a} and by \protect\cite{Riley2020} are added for comparison. Note that both contours were smoothed out for easier viewing. The colored full (lightly filled) circles in the left panel indicate the maximum masses of hyperonic (nucleon-only) stars.}
 \label{fig:6}
\end{figure}

In the multimesenger era, when not only data on masses and radii of NS are accumulating and becoming more precise, but also the analysis of the GW signals is continuously providing additional constraints, the choice of the EoS of the NS interior narrows. We use all this information to investigate cold NS models built using the QMC-A, CMF and DD2Y-T EoS. First, the TOV equations (\cite{Tolman1939,Oppenheimer1939}) are solved to yield the gravitational mass and radius of a non-rotating, spherically symmetric cold NS. The computed gravitational masses as a function of radii (right panel) and of the central baryonic density (left panel) are shown Fig.~\ref{fig:6}.

Our results are compared with the most recent data  from observation. \cite{Cromartie2019} reported gravitation mass of the J0740+6620 millisecond pulsar, obtained combining data from NANOGrav and the Green Bank telescope, to be 2.14$^{\rm +0.20}_{\rm -0.18}$ M$_\odot$ with 95.4\% credibility interval and 2.14$^{\rm +0.10}_{\rm -0.09}$ M$_\odot$ with 68.3\% credibility interval. \cite{Antoniadis:2013pzd} studied the PSR J0348+0432 neutron star-white dwarf systems and derived the pulsar mass as 2.01$\pm$0.04 M$_\odot$.

 Further observation of the binary millisecond pulsar PSR J1614-2230, reduced the original results, 1.97$\pm$0.04 M$_\odot$ to 1.928$\pm$0.017 M$_\odot$ (\cite{Fonseca2016}) and further to 1.908$^{\rm +0.016}_{\rm -0.0016}$ M$_\odot$ (\cite{Arzoumanian2018a}) with 68\% credibility interval.  Complementary to those measurements, \cite{Rezzolla2018}, combining the GW observations of BNSM and quasi-universal relations, set constraints on the maximum-mass configuration that can be attained by non-rotating NS.  The study yielded limits on the maximum mass of a non-rotating NS between $2.01^{\rm +0.04}_{\rm -0.04}$ and $2.16^{\rm +0.17}_{\rm -0.15}$ M$_\odot$. The QMC-A, CMF and DD2Y-T models with hyperons produce NS with the maximum gravitational mass consistent within these limits. The maximum mass of a purely nucleonic NS lies within the observational limits for the QMC-A and CMF EoS, but is somewhat higher for the DD2Y-T EoS.

The deduction of stellar radii for the maximal mass configuration from observation is rather involved. There are many estimates in the literature (see e.g. \cite{Oezel2016}), but the constraints they provide are still rather wide, 10 - 15 km,  as a simultaneous observation of a heavy NS and its radius has not been yet achieved. The much needed information for constraining the theory is complicated because the TOV equation yields only the  gravitational mass of a NS as a function of its radius. Recent GW observation directed the attention to lower-mass stars, with masses around the canonical value of 1.4 M$_\odot$. However, the GW data supply only information on masses and tidal deformation. Constraints on radii have to be inferred, often in combination with data from electromagnetic observation, in a model dependent way (\cite{Abbott2018,Raithel2019,Weih2019}).  Note that even when using universal relations to obtain stellar radii (\cite{Yagi:2013bca,Chatziioannou2018,Abbott2018}, the equation of state dependency cannot be fully eliminated.

\cite{Steiner2013} employed data on transiently accreting and bursting low mass X-ray binaries sources and  obtained  limits on the radius of a 1.4 M$_\odot$ star, R$_{\rm 1.4}$, between 10.4 and 12.9 km, independent of the structure of the core.  The upper limit on R$_{\rm 1.4}$  was found to be 13.6 km by \cite{Annala2018}, who used a piece-wise polytropic EoS, compatible at high densities with pQCD and using the GW limits on tidal deformability. \cite{Burgio2019} translated the limits on average tidal deformability, imposed by the GW signal, into limits on the R$_{\rm 1.5}$ (R$_{\rm 1.4}$) to be 11.8 (13.1) km. \cite{Raithel2019} obtained limits on the R$_{\rm 1.4}$ to be 9.8 - 13.2 km from Bayesian analysis of the GW data. Recently, \cite{Capano2020} constructed a large number of EoS based on the effective field theory and marginalized them using the GW observations. They obtained R$_{\rm 1.4}$=11.0$^{\rm +0.9}_{\rm -0.6}$ km (90\% credibility interval), with the upper limit lower than the one of \cite{Burgio2019}. Very recently, \cite{Al-Mamun2020} combined electromagnetic and gravitational wave constraints in a Bayesian analysis and obtained a set of 1$\sigma$ and 2$\sigma$ constraints on R$_{\rm 1.4}$ with mean points around 12 km.  But note that all these limits are, at least to certain extend, dependent on the EoS used in the analysis. This dependence is apparent not only in comparison of results of different models, but even within one model, considering different nuclear interactions, as was recently shown by \cite{Dexheimer:2018dhb}) in their study of the radius-tidal deformation relation.

\begin{figure}
  \includegraphics[trim={0 2.5cm 0 1.0cm},width=9.0cm]{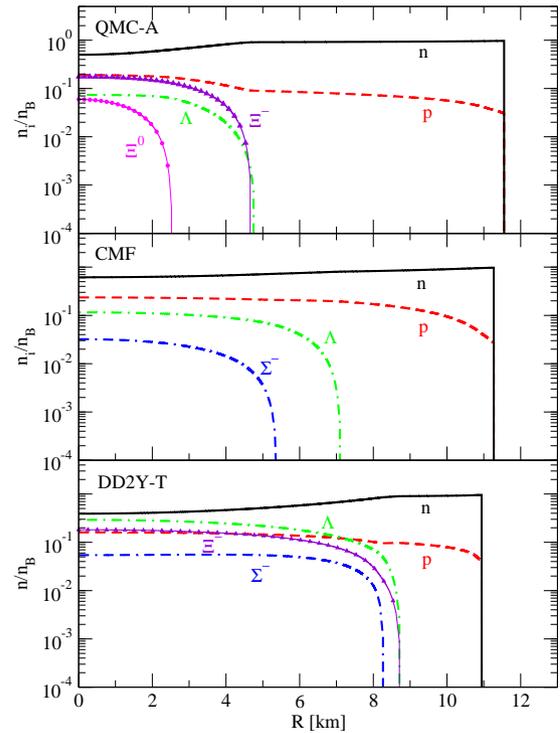}
  \caption{ Radial distribution of the nucleon and hyperon population in a maximum-mass NS at T=0 MeV.  Only population fractions higher than 10$^{-\rm 4}$ are shown.}
 \label{fig:7}
\end{figure}

The only observational data known to us, which report gravitational mass and the corresponding radius on the same object, are the results from the NICER mission. Bayesian inference approach of the energy-dependent thermal X-ray waveform of the isolated 205.53 Hz millisecond pulsar PSR J0030+0451 yields its estimated mass 1.44$^{\rm +0.15}_{\rm -0.14}$ M$_\odot$ and the equatorial circumferential  radius 13.02$^{\rm +1.24}_{\rm -1.06}$ km with 68\% confidence level \cite{Miller2019}, consistent with the outcome of an independent analysis by \cite{Riley2019}  1.34$^{\rm +0.15}_{\rm -0.16}$ M$_\odot$ and radius 12.71$^{\rm +1.14}_{\rm -1.19}$ km at the same credible interval. We display M-R correlation obtained by \cite{Miller2019a} (for the case of two potentially overlapping ovals) and by \cite{Riley2020} (run1) in the right panel of Fig.~\ref{fig:6}. We refer the reader for more details to these references.

\cite{Lattimer2005} proposed that EoS-independent Tolman VII solution to Einstein's equations sets an upper limit to the central density of  cold, non-rotating NS with maximum-mass configurations between 1.8 - 2.1 M $_\odot$ in the range 9 - 10 n$_{\rm B}$/n$_{\rm 0}$ (see Fig. 1 in \cite{Lattimer2005}). All three models in this work are well within this limit with central densities 4.75, 5.69 and 5.87 n$_{\rm B}$/n$_{\rm 0}$ for QMC-A, CMF and DD2Y-T models, respectively, as demonstrated in the left panel of Fig.~\ref{fig:6}.  It is interesting to note that this central density seems to be rather independent of the hyperonic core make up. 

 It is instructive to examine density distributions of nucleons and hyperons in cold maximum-mass NS configurations. As illustrated in Fig.~\ref{fig:7}, the distribution shows the expected sensitivity to a model EoS. All hyperon species disappear at roughly 5, 7, and 9 km from the stellar center in the QMC-A, CMF and DD2Y-T models, respectively.  Beyond these thresholds, the NS matter is composed only of nucleons and leptons (not shown here). 

\begin{figure}
  \includegraphics[trim={0 2.5cm 0 1.0cm},width=9.0cm]{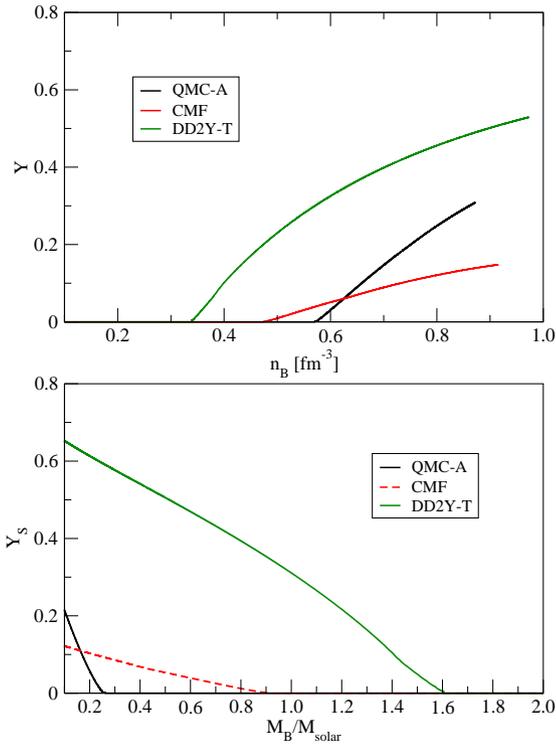}
  \caption{Total hyperon fraction vs. baryon number density in a maximum-mass star as calculated in the QMC-A, CMF and DD2Y-T models at T=0 MeV (top panel) and the total strangeness fraction in the enclosed baryonic mass of the NS (bottom panel).}
 \label{fig:8}
\end{figure}

The hyperon content in the PNS can be further quantified by examination of the total hyperon fraction $Y=\sum_{\rm j} n(j)/n_{\rm B}$ and total strangeness fraction $Y_{\rm s}=\sum_{\rm j} S(j) n(j)/n_{\rm B}$, where $j$ denotes a hyperonic species and $S$ the strangeness number. The baryon number density dependence of the total hyperon fraction (top panel) and the fraction of strange matter in a star with a given baryon mass (bottom panel) are illustrated in Fig.~\ref{fig:8}. A significant difference in total hyperon populations in the three models, the QMC and the CMF, when compared to the DD2Y-T model, is apparent. In particular, the DD2Y-T model predicts a  larger amount of hyperons at high densities, as compared to the QMC-A and CMF models. The large increase in the strangeness population reflects mainly in the lowering of the maximum mass of cold NS in the DD2Y-T model. But also other differences in, for example, as stellar radii and the adiabatic index, can be traced back to the strangeness distribution.

\subsubsection{Neutron stars at finite temperatures (PNS)}
\label{sec:ns_hot}

\begin{figure}
  \includegraphics[trim={0 2.5cm 0 1.0cm},width=9.0cm]{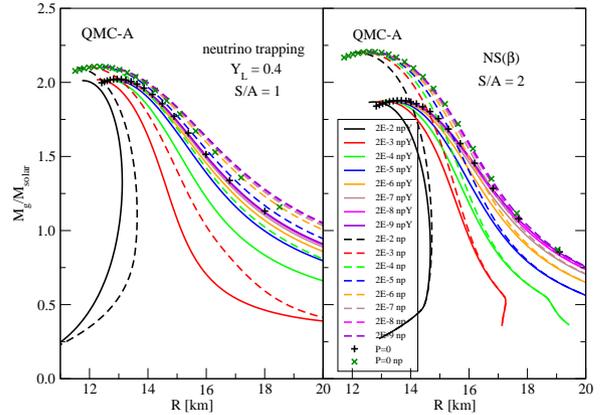}
  \caption{Illustration of the search for the surface of a hot star in the PNS-I  (left) and the PNS-II (right) cases,  as performed in the QMC-A model for stars with nucleons and hyperons in the core (npY) and nucleons only (np). Only results for $n_{\rm B}$ in the region  2.0$\times10^{\rm -2}$ - 2.0$\times10^{\rm -9}$ fm$^{\rm -3}$ are shown for clarity.}
 \label{fig:9}
\end{figure}

Calculation of the structure of a hot PNS is complicated by the uncertainty in the location of the neutrinosphere, which is needed as a boundary condition for solution of the TOV equation. We have therefore examined the dependence of the maximum stellar mass and the corresponding radius by solving the TOV equations up to a fixed baryon number density in the region of 2.0$\times10^{\rm -2}$ - 2.0$\times10^{\rm -12}$ fm$^{\rm -3}$, instead of locating the surface of the star at zero pressure. The results, shown in Fig.~\ref{fig:9}, demonstrate that the maximum gravitational mass and radius, computed at the surface baryon number density lower than n$_{\rm B} \leq \sim$2.0$\times^{\rm -9}$ fm$^{\rm -3}$, are practically identical to those obtained with the definition of the surface at zero pressure. This conclusion holds for all scenarios and the EoS considered in this work.  We observed a minor difference in radii of the lower mass models in the case of matter with trapped neutrinos, which is most likely reflecting the accuracy of the calculation at a very low particle number density. The study was repeated for the CMF and DD2Y-T models with very similar results. Therefore we have adopted 2.0$\times10^{\rm -12}$ fm$^{\rm -3}$ as the surface density in all cases.

\begin{table}
\footnotesize
\setlength{\tabcolsep}{0pt}
\centering
\caption{\label{tab:2} Macroscopic properties of non-rotating stars, as predicted in the QMC-A, CMF and DD2Y-T models in different scenarios. Maximum gravitational mass M$_{\rm g}^{\rm max}$ in units of M$_\odot$ and related radius R in km, the central pressure P and energy density $\epsilon$ in MeV/fm$^{\rm 3}$, and the central baryonic density $n_{\rm B}$ in fm$^{\rm -3}$  for different stars, scenarios and compositions are given. Percentage differences, ${\rm \Delta M_{\rm g}^{\rm stat}}$ and ${\rm\Delta R_{\rm stat}}$, between M$_{\rm g}$ and R of cold NS and the PNS-I and PNS-II cases. A positive (negative) value indicates increase (decrease) from the corresponding cold NS value. For more discussion see text. }
\vspace{5pt}
\begin{tabular}{ p{4cm}p{1.5cm}p{1.5cm}p{1.5cm} }
 Model             &       QMC-A   &  CMF &	DD2Y-T \\ \hline
{\bf{NS hyperons}}\\					
M$_{\rm g}^{\rm max}$              	&  1.963  &  2.075  &  2.091 \\
R	                                &  12.42  &  12.04  & 11.47  \\
P                    &   210.7  & 337.4  & 435.1        \\
$\epsilon$          & 1005   &  1111   & 1221    \\
n$_{\rm B }$	       &  0.872 & 0.908  & 0.973  \\ \hline
{\bf{NS nucleons}}\\
M$_{\rm g}^{\rm max}$	               & 2.125   & 2.137   & 2.417   \\
R	                               & 11.37   & 12.16   & 11.87   \\
P                    & 502.8   &  352.3  &   519.4         \\
$\epsilon$          & 1273   & 1085  & 1099  \\
n$_{\rm B}$	        & 1.019	& 0.892  &  0.851 \\ 
R$_{\rm 1.4}$	                        & 13.55	& 14.14     &  13.17  \\
P$_{1.4}$              &  45.9    &   56.8     &  46.8   \\
$\epsilon_{\rm 1.4}$   & 387.7   & 329.0   &  353.3  \\
n$_{\rm 1.4}$	        & 0.388	& 0.328  & 0.353  \\ \hline
{\bf{PNS-I hyperons}}\\
M$_{\rm g}^{\rm max}$	                & 2.022  & 2.085    & 2.177  \\
R	                                & 12.99 & 12.51    &12.46 \\
P                     &   256.5  &   369.1 &  414.0       \\
$\epsilon$           &1122 & 1208  & 1143 \\
n$_{\rm B}$	        &0.880   & 0.937    & 0.883  \\
${\rm \Delta M_{\rm g}^{\rm stat}}$            &  3.0   &   0.48    &   4.0  \\
${\rm\Delta R_{\rm stat}}$                 &   4.5  &  3.8     &   8.3   \\   \hline         
{\bf{PNS-I nucleons}} \\
M$_g^{\rm max}$	                &2.107   & 2.041    & 2.372   \\
R                                	&12.40   & 12.69   &  12.30   \\
P                     & 410.0  &   376.7  &   590.6          \\
$\epsilon$           &1237    & 1197    & 1139  \\
n$_{\rm B}$	        &0.940   &0.929    & 0.844   \\ 
${\rm \Delta M_{\rm g}^{\rm stat}}$             &  -0.85   &   -4.6    &    -1.9   \\
${\rm\Delta R_{\rm stat}}$                &   8.7    &  4.3    &    3.6\\   \hline         
{\bf{PNS-II hyperons}}\\
M$_{\rm g}^{\rm max}$              	&  1.966   & 2.077   &   2.068 \\
R	                                &  13.61  &  12.57  &    12.08  \\
P                     &   181.8  &     317.6  & 428.9        \\
$\epsilon$           &     956.9    & 1075    & 1248  \\
n$_{\rm B }$	        &	0.815   &	0.862  &  0.963\\ 
 ${\rm \Delta M_{\rm g}^{\rm stat}}$            &  0.15   &   0.10    &   1.1   \\
${\rm\Delta R_{\rm stat}}$                &   9.1    &  1.2     &    5.2   \\   \hline         
{\bf{PNS-II nucleons}} \\
M$_g^{\rm max}$	                &   2.205   &  2.203   &  2.426  \\
R                                   	&  12.65   &   13.5   &   12.79    \\
P                    &   329.8    &  343.0  &   463.8        \\
$\epsilon$           &     1111   &    978.4 &   1039    \\
n$_{\rm B }$	        &	0.874   &	0.783  &	0.789 \\ 
${\rm \Delta M_{\rm g}^{\rm stat}}$            &  3.7   &   2.0    &    4.0   \\
${\rm\Delta R_{\rm stat}}$                 &   11   &  10    &    7.5   \\   \hline         
				
\end{tabular}
\end{table}

\begin{table}
\footnotesize
\setlength{\tabcolsep}{0pt}
\centering
\caption{\label{tab:3}Percentage differences in the maximum gravitational mass and its radius in non-rotating nucleonic and hyperonic stars as predicted by QMC-A, CMF and DD2Y-T models under the same conditions. A positive (negative) value indicates increase (decrease) due to hyperons from the corresponding nucleon-only NS and PNS values.}
\vspace{5pt}
\begin{tabular}{ p{2.5cm}p{2.0cm}p{2.0cm}p{2.0cm} }
 Model             &       QMC-A   &  CMF &	DD2Y-T \\ \hline
 {\bf{NS T=0}}  \\     
  M$_{\rm g}^{\rm max}$   &    -7.9          &   -2.9   &    -14.5     \\
  R                   &     8.8          &   -1.0   &       -3.4      \\
 {\bf{PNS-I}}  \\            
 M$_{\rm g}^{\rm max}$   &      -4.1         &  0.79   &     -8.6         \\
  R                  &       4.6         &   -1.4    &      1.3         \\
 {\bf{PNS-II}} \\            
 M$_{\rm g}^{\rm max}$    &    -11.5        &  -5.6     &     -15.9     \\
  R                   &     7.3         &   -7.1    &     -5.7       \\   \hline
\end{tabular}
\end{table}

The properties of hot non-rotating stars in the PNS-I and PNS-II scenarios, are summarized in Table~\ref{tab:2}, together with the same  properties of cold NS, added for comparison. To quantify the changes in the maximum gravitational mass and respective radius of cold stars in the PNS-I and PNS-II scenarios, we include the percentage differences, ${\rm \Delta M_{\rm g}^{\rm stat}}$ and ${\rm \Delta R_{\rm stat}}$, separately for nucleonic and hyperonic stars, to isolate thermal effects. 

 Starting with nucleon-only stars, we observe a slight decrease in maximum mass, as compared to cold stars,  of the order of 4.6\% (CMF) to 0.85\% (QMC-A)  in the lepton-rich stars with trapped neutrinos (PNS-I), indicating that the EoS is getting somewhat softer at high densities. The PNS-II scenario has the opposite effect on maximum mass, which increases between 4.0\% and 2.0\% (CMF). The radius of the star is increases by a maximum of  8.7\% (QMC-A) to 3.6\% (DD2) in the PNS-I and by between 11\% (QMC-A) and 7.5\% (DD2) in the PNS II. We see that the thermal effect and the effect of a difference in composition between a neutrino-trapped and chemically-equilibrated star on the maximum mass and the related radius are around ten percent or less in the nucleon-only stars for all the EoS models.

Turning to hyperonic stars, the maximum mass in the PNS-I case is marginally higher than in the cold ones, being between 4.0\% (DD2Y-T) and 0.48\% (CMF), indicating some stiffening of the EoS at high densities due the lower amount of hyperons at high densities in the lepton rich matter. The increase in the maximum gravitational mass of configurations with hyperons in the PNS-II case, as compared to cold stars, is almost negligible, being between 1.1\% (DD2Y-T) and 0.1\% (CMF). Similarly to nucleonic stars, the radius is increased in both scenarios, as compared to cold stars, by a maximum of 8.3\% (DD2Y-T) in the PNS-I and 9.1\% (QMC-A) in the PNS II. Again, all the changes vary at or below 10\%. Similarly to the nucleon-only stars, it is not possible to trace a consistent relation of the temperature and/or the difference between the PNS-I and PNS-II scenarios effects to a particular EoS model but there certainly is a relation between the amount of hyperons and the M-R curve.

The percentage difference in maximum gravitational mass and radius between nucleon-only and hyperonic stars, in the attempt to examine the effect of hyperons under the same thermodynamical conditions, is illustrated in Table~\ref{tab:3}. As expected, these results show some systematic sensitivity to a model EoS. In cold stars, the decrease in the maximum gravitational mass of a hyperonic star is 14.9\% (DD2Y-T) and 2.9\% (CMF), as compared to nucleon-only stars under the same conditions. Maximum mass of hyperonic lepton-rich stars in the PNS-I scenario lowers by between  8.6\% (DD2Y-T) and 4.1\%(QMC-A). It remains almost the same, with a 0.79\% increase in the CMF model, which is related to the fact that there is only small amount of hyperons present even when $Y_{\rm L}$ is not fixed. In the PNS-II scenario, the effect of hyperons is larger, as discussed in Sec.~\ref{sec:eos_hot}, lowering the maximum mass between 15.9\% (DD2Y-T) and 5.9\% (CMF). 

Interestingly, the QMC-A model predicts an increase in the radius of maximum-mass stars with hyperons in all cases, in contrast with the CMF and DD2Y-T models. We do not have an immediate explanation for this effect but, qualitatively, it is related to the amount of hyperons appearing at low densities. 

\begin{figure}
  \includegraphics[trim={0 2.3cm 0 1.0cm},width=9.0cm]{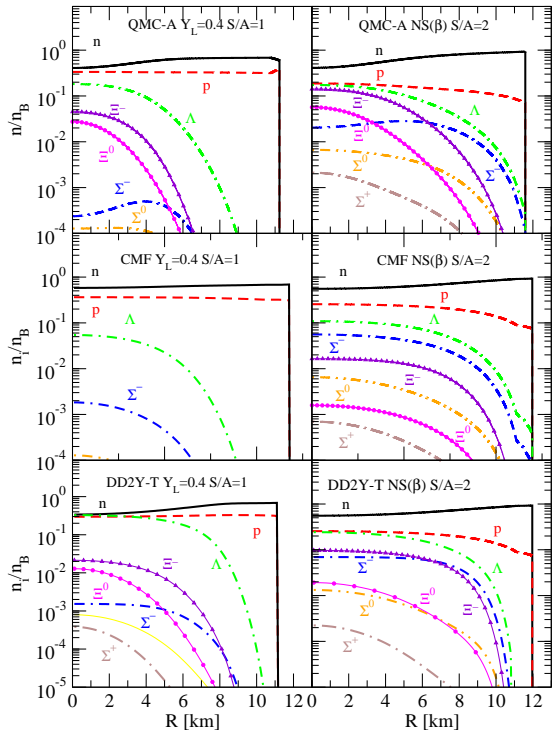}
  \caption{ The same as Fig.~\ref{fig:7} but for hot star in the PNS-I (left) and PNS-II (right) scenarios.}
 \label{fig:10}
\end{figure}

We have already discussed in Sec.~\ref{sec:eos_hot} that the population of hyperons grows with increasing density and temperature. In Fig.~\ref{fig:10} we demonstrate that radial distribution of the hyperon population inside the stellar core is also significantly impacted by increasing temperature. Comparing Fig.~\ref{fig:7} with Fig.~\ref{fig:10} shows that, for example, $\Lambda$ hyperons can be found almost throughout the whole volume of massive stars, reaching close to 10 km from the stellar center. This spreading is wider in the PNS-II case, in which the star is hotter than in PNS-I case (see Fig.~\ref{fig:4}). 

\begin{figure}
  \includegraphics[trim={0 1.0cm 0 1.0cm},width=9.0cm]{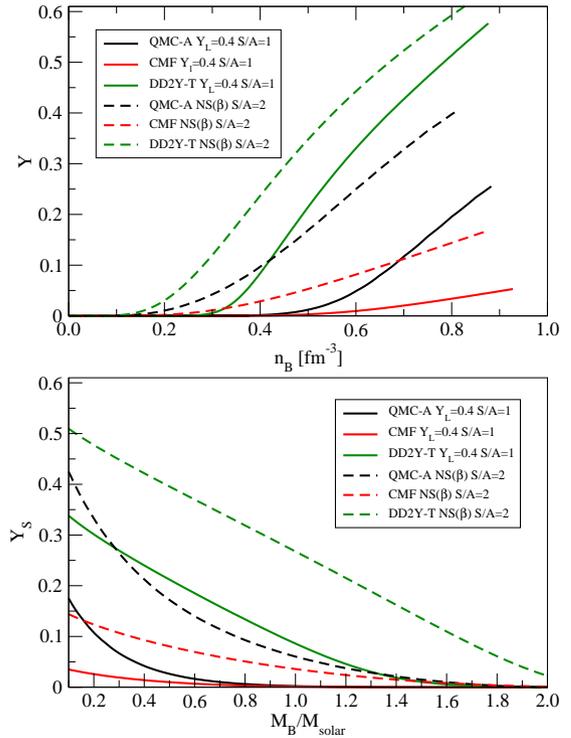}
  \caption{Total hyperon fraction vs. baryon number density as calculated in maximum-mass star as predicted by the  QMC-A, CMF and DD2Y-T models (top panel) and the total strangeness fraction in the enclosed neutron-star baryonic mass (bottom panel) at finite temperature in the PNS-I and PNS-II scenarios.}
 \label{fig:11}
\end{figure}

 In the same fashion, the total hyperon and strangeness fractions in the PNS-I and PNS-II scenarios, displayed in Fig.~\ref{fig:11}, indicate a substantial amount of strangeness present in a PNS just a few milliseconds after birth and increasing even more up to about 1 minute. This amount then decreases during the cooling process when some strangeness producing reactions freeze out and the composition of the NS core is fixed to what is expected in cold stars (see Fig.~\ref{fig:8}).

\section{Rotating stars}
\label{sec:rotation}

So far, we have investigated properties of cold and hot static NS. Now, we turn our attention to rigid rotation (differential rotation goes beyond the scope of the present manuscript), which deepens our discussion, as it affects not only the masses and radii of stars, but also their composition in a significant way. There is an observational evidence of fast rotating young millisecond pulsars (see e.g.\cite{Bassa2017a,Pleunis2017}).  Because of conservation of angular momentum, the PNS spin depends on the angular momentum of the progenitor star, a gain of angular momentum during the collapse \cite{Janka2016} and/or accretion of the downfall of material (\cite{Stockinger2020}). Interestingly, advanced CCSN simulations consider both rotating and non-rotating progenitors (\cite{Pajkos2019,Pan2020}). \cite{Bollig2020} reported a dramatic deceleration of the PNS outer layers with time up to 1.5 s after bounce. This prediction suggests the need to consider  differential rotation of outer and inner layers of a PNS in early times after bounce. In this work we adopt a rigid rotation model as the first step, leaving a more refined description of the PNS rotation for future work. 

Let us begin our discussion with the moment of inertia $I$, which is a candidate for providing a much needed observational constraint of the high-density EoS. It is proportional (using Newtonian physics for simplicity in this argument) to the mass $M$ of a star times its radius $R$ squared. Thus, if the mass of a star is known, its radius can be determined from observation of its moment of inertia. Together with the quadrupole moment, $Q$, and the Love number, $k_2$, reflecting the deviation from sphericity and the deformability of the star, $I$ is one of the global observables which is believed to exhibit universal relations (see \cite{Breu2016,Wei2019a} and references therein), which are approximately equation of state independent at zero temperature. Deviations from that universality, both for the moment of inertia as a function of compactness and gravitational mass of a star, and of the $I-Love-Q$ relation (not studied here)  at finite temperature have been suggested \cite{Martinon:2014uua,Marques2017,Lenka2019,Raduta2020}. 

\begin{figure}
  \includegraphics[trim={0 3.0cm 0 1.0cm},width=9.0cm]{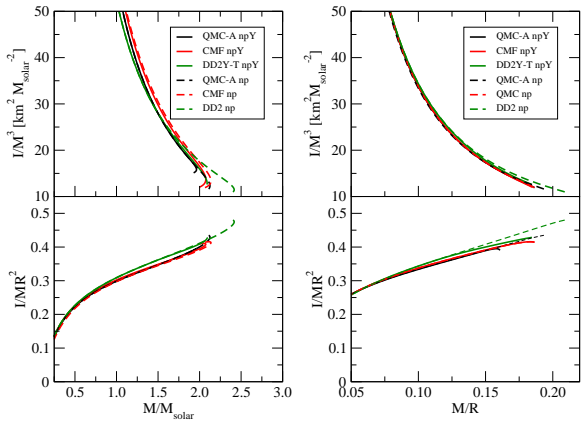}
  \caption{Normalized moment of inertia ${\rm I/M^{\rm 3}}$ (top) and ${\rm I/MR^{\rm 2}}$ (bottom) vs. the gravitational mass $M$ (left) and compactness $M/R$ (right) calculated for a nucleon-only star (np) and a star with nucleons and hyperons in the NS core (npY) at T=0 MeV. Note that the figure is organized at the same way as Figure 2 in (\protect\cite{Wei2019a}) to allow comparison with other EoS.}
 \label{fig:12}
\end{figure}

\begin{figure}
  \includegraphics[trim={0 2.5cm 0 1.0cm},width=9.0cm]{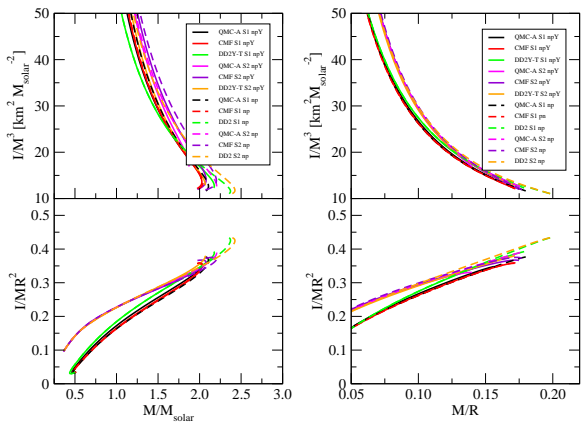}
  \caption{The same as Fig.~\ref{fig:12} but for a hot star in the PNS-I and PNS-II scenarios, labeled S1 and S2 in the figure, respectively, for the nucleon only (np) and nucleons and hyperons (npY) in the PNS core cases. Note that the axes are the same as in Fig.~\ref{fig:12} to easy comparison. }
 \label{fig:13}
\end{figure}

Here, we calculate two quantities, $I/M^3$, and $I/MR^2$, as a function of the maximum stellar gravitational mass and its compactness, $M/R$. The results for cold stars in Fig.~\ref{fig:12}, confirm that the QMC-A, CMF and DD2/DD2Y-T models for both nucleonic and hyperonic stars indeed exhibit very similar patterns for these quantities. Thus they are good candidates to be included in data sets leading to extraction of the most likely value of the stellar radius from a known mass and the moment of inertia. In contrast, for hot stars we see a difference between data for cold stars (see Fig.~\ref{fig:12}) as well as between the PNS-I and PNS-II scenarios, as illustrated in Fig.~\ref{fig:13}. Similar results were obtained by \cite{Raduta2020} who studied, among other effects, the relation between normalized moments of inertia and the compactness of a PNS in different thermodynamical situations for a number of EoS models.  Our results confirm the findings of \cite{Lenka2019}, who studied rapidly rotating massive stars with either nucleon
or the entire baryon compositions, both at zero and finite temperature. We conclude, in accord with \cite{Lenka2019}, that investigation of a temperature dependence of dynamical properties of NS may become an important input in post-BNSM simulations and deserves further investigation. Furthermore, recent data on the GW190425 event with one component having gravitational mass 2.0$^{\rm +0.6}_{\rm -0.3}$ M$\odot$ \cite{Abbott2020a,Abbott2020b} were analysed considering the spin of merging stars. The high mass of the component would also require EoS taking into account the hyperon content in the NS core, not necessarily needed when both merging NS have low mass, as in the case of GW170817.

The relation between the moment of inertia and the mass parameters of NS and PNS was explored  using a slow rotation code based on the Hartle-Thorne method (\cite{Miller2020}) It is, however, also interesting to explore fast rotating stars close to and at their Kepler frequency, the maximum frequency at which stars are still compact, not shedding matter. In particular, the limit on the maximum mass of a fast spinning star is interesting for the ongoing discussion of the possible identification of the secondary component with gravitational mass around 2.6 M$_\odot$ of GW190814 as a neutron star or a low mass black hole \cite{Abbott2020}). This observation stimulated a lot of activity in the community. The secondary object was interpreted as a fast spinning pulsar (see e.g. \cite{Dexheimer2020,Zhang2020,Biswas2020,Demircik2020}), a star with asymmetric core \cite{Roupas2020} or a product of a complicated mass transfer during the merger \cite{Safarzadeh2020}. Less confident results were reported from statistical analysis \cite{Godzieba2020} and the neutron star interpretation was rejected in, for e.g. \cite{Fattoyev2020,Lim2020}. Here we explore the upper limit on the gravitational mass of fast spinning cold NS, with possible relevance to the GW190814 and to stars in the PNS-I and PNS-II scenarios, which may be interesting for CCSN simulations.

\begin{figure}
  \includegraphics[trim={0 2.5cm 0 1.0cm},width=9.0cm]{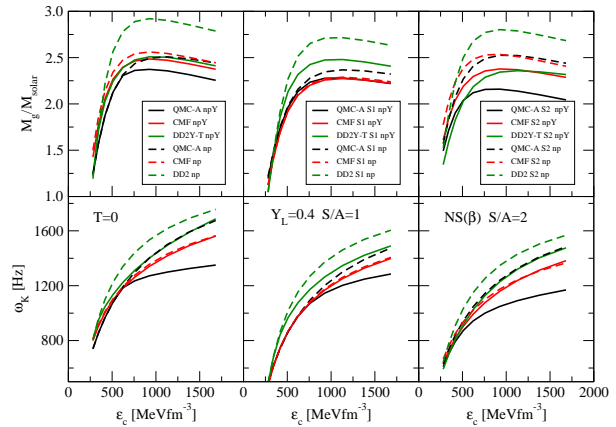}
  \caption{ Gravitational masses of NS and PNS equilibrium sequences rotating at Kepler frequency  (top) and the corresponding frequencies (bottom) vs. the central energy density. Results for cold stars (T=0) and hot stars in the PNS-I and PNS-II scenarios, labeled S1 and S2 in the top panels) are displayed in the left, middle and right panels, respectively. Data for nucleon-only (np) and the full baryon octet (npY), calculated in the QMC-A, CMF and DD2/DD2Y-T models, are shown.}
 \label{fig:14}
\end{figure}

In Fig.~\ref{fig:14}, we illustrate properties of stars rotating at Kepler frequencies as a function of central energy density as calculated in the QMC-A, CMF and DD2/DD2Y-T models using the publically available RNS code. Macroscopic properties of a star with maximum gravitational mass M$_{\rm g}^{max}$ in each case are given in Table~\ref{tab:4}, including equatorial radius  $R_{\rm e}$, ratio of the polar to equatorial axes $r_{\rm p}/r_{\rm e}$  and the ratio of the rotational to gravitational energies $T/W$ (\cite{Stergioulas1995,Stergioulas2003,Paschalidis2017}). The table is in the same format as Table~\ref{tab:2} for easier comparison with some properties of non-rotating stars. Percentage differences $\Delta M_{\rm g}^{rot}$ and $\Delta R^{rot}$ are added to illustrate the effect of the maximum rotation on the $M_{\rm g}^{max}$ and radius of the equivalent non-rotating stars. The effects on radius should be taken with care. Here we show only the equatorial radius $R_e$ as calculated in the RNS code. The uncertainty in determination of the oblate rotating star radii (adding to complications caused by thermal effects) has been recently discussed by \cite{Silva2020} and will follow addressed in a later publication.

\begin{table}
\footnotesize
\setlength{\tabcolsep}{0pt}
\centering
\caption{\label{tab:4} Macroscopic properties of stars, rotating at Kepler frequency, as predicted in the QMC-A, CMF and DD2Y-T models.  Maximum gravitational mass $M_{\rm g}^{\rm max}$ in units of  $M_\odot$, central energy density $\epsilon_{\rm c}$  in MeV/fm$^{3}$, radius at equator $R_e$ in km, Kepler frequency ${\rm \omega_{\rm K}}$ in Hz, the ratio of rotational kinetic and gravitational energies T/W and the polar to equatorial axis ratio $r_{\rm p}/r_{\rm e}$ are presented. Percentage differences ${\rm \Delta M_{\rm g}^{\rm rot}}$ and ${\rm\Delta R_{\rm rot}}$ are added for comparison. A positive (negative) value indicates increase (decrease) from the equivalent non-rotating stars (see Table~\ref{tab:2}).  }
\vspace{5pt}
\begin{tabular}{ p{4cm}p{1.5cm}p{1.5cm}p{1.5cm} }
 Model             &       QMC-A   &  CMF &	DD2Y-T \\ \hline
{\bf{NS hyperons}}\\					
 M$_{\rm g}^{\rm max}$                       	&  2.384  &  2.490  &  2.512 \\
$\epsilon_{\rm c}$ 	                &  903.8  &  955.7 & 1008  \\
 R$_{\rm e}$                             &   16.84  & 16.40 & 15.73         \\
${\rm \omega_{\rm K}}$                   & 1280   &  1354   & 1444    \\
T/W   	                       &  0.116 & 0.119 & 0.121  \\ 
r$_{\rm p}$/r$_{\rm e}$                     & 0.352  & 0.334   &  0.314 \\
${\rm \Delta M_{\rm g}^{\rm rot}}$       &  19.4      & 18.2 & 18.3   \\
${\rm\Delta R_{\rm rot}}$        &  30.2      &  30.7 & 31.3  \\  \hline
{\bf{NS nucleons}}\\
M$_{\rm g}^{\rm max}$                       	&  2.517  &  2.565   &  2.920 \\
$\epsilon_{\rm c}$  	                &  1085  &  955.7  & 913.6 \\
 R$_{\rm e}$                             &   15.38  & 16.43 &   15.85  \\
${\rm \omega_{\rm K}}$                  & 1491  &  1369  & 1528     \\
T/W   	                       &  0.115 & 0.118 & 0.135\\ 
r$_{\rm p}$/r$_{\rm e}$                    & 0.304  & 0.331 & 0.295 \\
${\rm \Delta M_{\rm g}^{\rm rot}}$            &   16.9 &  18.2 & 18.9 \\
${\rm\Delta R_{\rm rot}}$             &    30.0 & 29.9   & 28.7 \\  \hline
{\bf{PNS-I hyperons}}\\
M$_{\rm g}^{\rm max}$                       	& 2.282  & 2.278  &  2.484 \\
$\epsilon_{\rm c}$  	                &  1046 &  1125  & 1050 \\
 R$_{\rm e}$                             &   17.41 & 16.66 & 16.58     \\
${\rm \omega_{\rm K}}$                   & 1182&  1258  & 1319 \\
T/W   	                       &  0.0785 & 0.0807  & 0.0915 \\ 
r$_{\rm p}$/r$_{\rm e}$                    &  0.375  &  0.354 & 0.341 \\
${\rm \Delta M_{\rm g}^{\rm rot}}$           &  12.1  &  11.8  & 13.2 \\
${\rm\Delta R_{\rm rot}}$            &  29.1  &  28.5  & 28.4  \\  \hline
{\bf{PNS-I nucleons}} \\
M$_{\rm g}^{\rm max}$                       	&  2.368  &  2.293  &  2.710\\
$\epsilon_{\rm c}$  	                &  1167  &  1130 & 1058  \\
 R$_{\rm e}$                             &   16.35  & 16.60 & 16.31       \\
${\rm \omega_{\rm K}}$                   & 1315  &  1269  & 1418   \\
T/W   	                       &  0.0814& 0.0813  & 0.101\\ 
r$_{\rm p}$/r$_{\rm e}$                     &  0.341  &   0.351 & 0.314 \\
${\rm \Delta M_{\rm g}^{\rm rot}}$            & 11.7   &  11.6  & 13.3  \\
${\rm\Delta R_{\rm rot}}$            &  27.5  &  26.7  &  28.0  \\ \hline
{\bf{PNS-II hyperons}}\\
M$_{\rm g}^{\rm max}$                       	&  2.161  & 2.378  &  2.241\\
$\epsilon_{\rm c}$  	                &  871.3  &  937.5 & 1126 \\
 R$_{\rm e}$                             &   18.79 & 17.85  & 17.90       \\
${\rm \omega_{\rm K}}$                   & 1035 &  1165  & 1139   \\
T/W   	                       &  0.0807 & 0.0860  & 0.0859 \\ 
r$_p$/r$_e$                    &  0.418   & 0.382 & 0.390\\
${\rm \Delta M_{\rm g}^{\rm rot}}$           &  9.45     & 13.51  & 8.03  \\
${\rm\Delta R_{\rm rot}}$            & 32.0   & 34.7  & 38.8 \\  \hline
{\bf{PNS-II nucleons}} \\
M$_{\rm g}^{\rm max}$                       	&  2.527  &  2.534  &  2.788 \\
$\epsilon_{\rm c}$  	                &  972.37 &  871.3 & 937.5\\
 R$_{\rm e}$                             &   17.21  & 18.35  & 16.91      \\
${\rm \omega_{\rm K}}$                   & 1262   &  1182   & 1353   \\
T/W   	                       &  0.0918 & 0.0906  &  0.104 \\ 
r$_{\rm p}$/r$_{\rm e}$                     & 0.357  &  0.388 & 0.332\\ 
${\rm \Delta M_{\rm g}^{\rm rot}}$             & 13.6  &  14.0 & 13.9 \\	
${\rm\Delta R_{\rm rot}}$             & 30.5   &  30.5 &  27.7 \\  \hline			
\end{tabular}
\end{table}

\begin{table}
\footnotesize
\setlength{\tabcolsep}{0pt}
\centering
\caption{\label{tab:5}Percentage differences in the maximum gravitational mass, its radius and the Kepler frequency of cold and hot nucleonic and hyperonic stars, as predicted by QMC-A, CMF and DD2/DD2Y-T models. A positive (negative) value indicates increase (decrease) due to hyperons from the corresponding nucleon-only star value .}
\vspace{5pt}
\begin{tabular}{ p{2.5cm}p{2.0cm}p{2.0cm}p{2.0cm} }
 Model             &       QMC-A   &  CMF &	DD2Y-T \\ \hline
 {\bf{NS T=0}}  \\     
  M$_{\rm g}^{\rm max}$   &     -5.3          &   -3.0   &    -15     \\
  R                   &     9.1          &   -0.18   &    -0.76      \\
  $\omega_{\rm K}$  &    -15         &     -8.5    &     -15       \\
 {\bf{PNS-I}}  \\            
 M$_{\rm g}^{\rm max}$   &      -3.7         &  -0.65   &     -8.7         \\
  R                  &      6.3         &   0.36    &     1.6         \\
 $\omega_{\rm K}$  &     - 11         &   -8.5   &    -7.2        \\  
 {\bf{PNS-II}} \\            
 M$_{\rm g}^{\rm max}$    &    -15        &  -6.4     &     -16     \\
  R                   &     8.8         &   -2.3    &     5.7       \\
 $\omega_{\rm K}$   &    -19         &  -1.5     &    -17      \\  \hline
\end{tabular}
\end{table}

Following the same pattern used in our discussion of properties of non-rotating stars, we first compare cold and warm stars, rotating at Kepler frequency, with the same particle make-up.  Looking at the nucleon-only cold NS, the rotation increases the maximum mass by a very similar amount between 18.9\% (DD2Y-T) and 16.9\% (QMC-A). The increase is smaller for the  PNS-I and PNS-II scenarios, varying only slightly between 11 - 14\% in both scenarios and all the three EoS models. A similar pattern is found for the increase in the equatorial radius, being around 30\% in all cases.

 In cold hyperonic stars, the maximum mass increase is very similar to that in the nucleon only configuration, from 19.4\% (QMC-A) to 18.2\% (CMF). This increase is lower both in the PNS-I and PNS-II scenarios, being between 13.2\% (DD2Y-T) and 11.6\% (CMF) in PNS I and even lower in the PNS-II 13.5, 9.45 and 8.03\%  in CMF, QMC-A and DD2Y-T, respectively. This trend indicates that the increase in the maximum mass due to the rotational energy competes with the decrease due to the growing amount of hyperons in the PNS-II scenario. It is interesting to see that in the PNS-II case the increase in the equatorial radius is anti-correlated with the maximum mass, being largest (38\%) with the smallest maximum mass increase predicted by the DD2Y-T EoS. 

A comparison between nucleon-only and hyperonic stars under the same thermodynamic conditions is illustrated in Table~\ref{tab:5}. Examination of the table shows that the hyperons lower the maximum gravitational mass, similarly to the non-rotating stars, in all models. This effect is strongest in the scenario PSN II, with the highest hyperon content, as expected. We did not find a consistent trend in the $R_{\rm e}$ radius. The QMC-A model predicts increase by up to 10\% in all scenarios. The change in $R_{\rm e}$ varies by less than 5\% in either direction. As noted before, the value of R$_{\rm e}$ may be a rather schematic indicator of shape changes in the RNS simulation.

The upper limit on the gravitational mass of maximally rotating cold neutron stars (see the top two sections of Table~\ref{tab:4}) is close or above the expected mass of the secondary GW190425. Therefore, we cannot exclude a possibility that the secondary is a heavy, fast rotating, pulsar. The masses of hot stars (lower four sections of Table~\ref{tab:4}) are somewhat lower, thus supporting the notion that the secondary is cold. Nevertheless, if the secondary were made of nucleons only, this argument would be weaker.
  
 Finally, all models predict lower Kepler frequency for hyperonic stars than the nucleonic stars under the same conditions. Thermal effects also lower the Kepler frequency because the hot stars, being less compact, reach the mass-shedding limit at lower angular frequency. These effects are reflected in the ratio of the rotational and gravitational energy ratio $T/W$. The typical value of this ratio is calculated to be about 0.11 in cold stars and about 0.09 in warm stars. The effect of hyperons is however slight and the values are similar in all models.

\section{Summary and Outlook} 
\label{sec:discussion}

In this work, we presented for the first time results of the QMC-A model, based on sub-baryon degrees of freedom and having only five, well constrained variable parameters. Three different stages of stellar evolution were considered, the lepton rich PNS with trapped neutrinos (PNS-I), the neutrinoless chemically equilibrated PNS (PNS-II) and the cold catalyzed NS, the final stage of the evolution. In all cases, a substantial amount of hyperons was found in the core of massive stars, larger in the first two evolution stages than in the later one, signaling a dependence of the presence of strangeness on the thermodynamical environment. Comparing the first two stages, we confirm, in line with some other studies, that the large electron content inhibits the hyperon degrees of freedom in PNS with trapped neutrinos. In the neutrinoless, deleptonized PNS, the hyperon content increases in order to keep the charge neutrality of the matter.

When uniform slow rotation in the Hartle-Thorne approximation is included in the calculation, our moment of inertia relations to the gravitational mass and compactness of a maximum-mass star show the expected universal relations for cold NS but breaks the relations for hot PNS. To contribute to the discussion on the nature of of the second component of the GW190814 event, we calculated upper limits on the maximum gravitational mass of cold, relevant for GW190814, and hot nucleonic and hyperonic stars, when the rotation frequencies were taken to the mass-shedding Kepler limit, using the publicly available RNS code.

In this work, we have also illustrated the effect of the appearance of hyperons on the adiabatic index and on the speed of sound $c_{\rm s}$. We found that a low speed of sound within the conformal limit at large densities $c^{\rm 2}_{\rm s} <1/3$ can be reproduced by our hadronic model, not being in this case a fingerprint of quark matter cores in neutron stars, but a consequence of instabilities caused by the onset of hyperons. Our results re-open the question of the existence of r-modes in rotating neutron stars (\cite{Andersson2001,Haskell2015}). \cite{Jones2001,Jones2001a} reported that the bulk viscosity of hyperonic matter in neutron stars would produce a serious damping of the r-modes.  Lidblom and Owen \cite{Lindblom2002} argued that although the cooling of the PNS is too rapid to influence the r-modes. Very recently \cite{Zhou2020}  studied r-mode stability of the secondary component of GW190814 provided it is interpreted as supermassive and superfast pulsar. It will be interesting to pursue the connection between r-modes and the internal composition of neutron stars in the future.  

Throughout this study, we compared the QMC-A models with results of two other models, the chiral mean-field model (CMF), and the DD2/DD2Y-T, employing the generalized relativistic density functional with hyperons (GRDF-Y). We have made a systematic comparison of a wide range of observables of cold and hot nucleon-only and hyperonic stars in the three models, based on very different physics, and found that they all satisfy basic observational and empirical constraints on dense matter in neutron stars. In particular, the maximum gravitational masses of both cold and hot hyperonic stars agree with observation within current limits, and do not show any sign of the often discussed ``hyperon puzzle''. But to give preference and make positive distinction between the models, additional data, more sensitive to microphysics, would be needed. Future data from NICER and analysis of BNSM with at least one component having a high mass, such as GW190425, may offer such information.   

As a final product of this work, we have constructed QMC-A EoS tables, containing data in the parameter space compatible with CCSN and BNSM simulations. These tables will be posted on the CompOSE depository (http://compose.obspm.fr)  in the near future.

\section*{Acknowledgments}

We are indebted to John Miller for suppling some of the essential software for analysis of the model data and many elucidating discussions. Raph Hix and Bronson Messer are acknowledged for elucidating discussions related to  CCSN events.  JRS and PAMG acknowledge a fruitful discussion with Andrew Steiner during the course of writing the computer code used in this work and hospitality during their stay at the University of Adelaide.
This work was supported by the University of Adelaide and the Australian Research Council through the ARC Centre of Excellence in Particle Physics at the Terascale 
(CE110001004) and grants DP180100497 and DP150103101. VD is supported by the National Science Foundation under grant PHY-1748621 and PHAROS (COST Action CA16214).

\section*{Data availability}
The data underlying this article will be shared on reasonable request to the corresponding author.

\bibliographystyle{mnras}
\bibliography{QMC_Sept_2020}
\end{document}